\documentclass[aps,prl,reprint,superscriptaddress,amsmath,amssymb,floatfix,showpacs]{revtex4-1}
\usepackage[T1]{fontenc}
\usepackage{mathtools}
\usepackage{bm}
\usepackage{amsmath}
\usepackage{graphicx}
\usepackage{esint}
\usepackage{float}
\usepackage{subfig}
\makeatletter

\usepackage{tikz}
\usetikzlibrary{calc}

\usepackage{color}
\usepackage{epstopdf}
\usepackage{slashed}
\usepackage{bm}
\usepackage{braket}
\usepackage{soul}
\usepackage{ulem}

\makeatother

\begin{document}

\title{Nematicity in iron pnictides: phase competition and emergent symmetry}
\author{Yiming Wang}
\affiliation{Department of Physics and Astronomy, Extreme Quantum Materials Alliance, Smalley-Curl Institute, 
Rice University, Houston, Texas 77005, USA }
\affiliation{Department of Physics and Beijing Key Laboratory of Opto-electronic
Functional Materials and Micro-nano Devices, Renmin University of
China, Beijing 100872, China }

\author{Changle Liu}
\affiliation{School of Physics and Mechatronic Engineering, Guizhou Minzu University, Guiyang 550025, China }

\author{Shan Wu}
\affiliation{Department of Physics, University of California at Berkeley, Berkeley, California 94720, USA}
\affiliation{Department of Physics and Engineering Physics, Santa Clara University, Santa Clara, California, 95053, USA}
\author{Jianda Wu}
\email{wujd@sjtu.edu.cn}
\affiliation{Tsung-Dao Lee Institute \& School of Physics and Astronomy, Shanghai
Jiao Tong University, Shanghai 200240, China}

\author{Qimiao Si}
\email{qmsi@rice.edu}
\affiliation{Department of Physics and Astronomy, Extreme Quantum Materials Alliance, Smalley-Curl Institute, 
Rice University, Houston, Texas 77005, USA }

\author{Rong Yu}
\email{rong.yu@ruc.edu.cn}
\affiliation{Department of Physics and Beijing Key Laboratory of Opto-electronic
Functional Materials and Micro-nano Devices, Renmin University of
China, Beijing 100872, China }
\affiliation{Tsung-Dao Lee Institute \& School of Physics and Astronomy, Shanghai
Jiao Tong University, Shanghai 200240, China}

\begin{abstract}
The phase diagram of iron-based superconductors contains 
a host of electronic orders, which are intimately connected with their 
superconductivity.
Here we analyze the fluctuations of one type of nematic order in another.
Our analysis leads to an emergent U(1) symmetry at 
a first-order transition between a nematic phase and a $C_4$-symmetric charge-ordered phase. 
We characterize
the continuous symmetry in terms of a certain hidden Lie algebra that links the different orders. 
This emergent 
symmetry leads to a Goldstone mode at the transition and causes softening of excitations in the nematic and charge sectors
near the transition. 
The underlying physics bears a resemblance to the anisotropic XZ spin model, 
with the nematic order and charge $C_4$ order parameters playing the roles of the $x$ and $z$ components of the magnetization vector,
respectively.
We provide
the experimental 
evidence
in support of the proposed effects, and discuss the general implications of our results for the physics of iron-based superconductors 
and other correlated systems.
\end{abstract}

\maketitle
\noindent
\textit{Introduction.~} Understanding complex phases
is one major theme of condensed matter physics~\cite{Kei17.1,Pas21.1,Fradkin_RMP:2015}. 
At low temperatures, phases are usually characterized by order parameters that describe
spontaneous symmetry breaking. Quantum fluctuations 
are amplified near quantum phase transitions. 
It has been widely recognized that,
near a 
quantum phase transition, new phases and exotic properties may be nucleated,
such
as quantum spin liquids, 
strange metallicity
and unconventional superconductivity
~\cite{Balents_Nature:2010,Takagi_NRP:2019,Kir20.1,Lee_RMP:2006}.
Another increasingly appreciated feature is that, 
near a quantum phase transition,
additional symmetries may emerge
that are not present in
the underlying microscopic Hamiltonian.
For example, 
an emergent SO(5) symmetry was 
proposed 
in the context of cuprate superconductivity~\cite{Zhang_Science:1997}.
Related behavior has been 
discussed in a variety of 
contexts~\cite{Grover_Science:2014, Yao_PRL:2017, Yao_SA:2018, Moessner_PRB:2001, Balents_PRB:2002, Hermele_PRB:2004, Moessner_PRB:2003, Blankschtein_PRB:1984, Coldea_Science:2010, Wu_PRL:2021,Tanaka_PRL:2005,Nahum_PRL:2015}. 
One common mechanism for the emergent symmetry lies in the entwining of two distinct orders: 
at the phase transition, 
this symmetry
can rotate the two orders into each other. 

In iron-based superconductors, 
superconductivity is accompanied by various
electronic orders. In addition to 
the antiferromagnetic (AFM) order, these include 
various forms of nematic orders~\cite{Kamihara_JACS:2008, Johnston_AP:2010, Dai_RMP:2015, Si_NatRevMat:2016, Hirschfeld_CRP:2016, Wang_science:2011, MYi:2011, IFisher:2012,Si-Hussey2023}.
The AFM state 
is ordered at the wave vector
$(\pi,0)$ or $(0,\pi)$ and reduces the lattice symmetry 
from tetragonal ($C_4$ rotational) to orthorhombic ($C_{2}$). 
The nematic order breaks  the lattice rotational symmetry, and is often 
understood as a composite
order associated with the anisotropic AFM 
and/or
other, related,
fluctuations within
the spin-driven
description~\cite{FangKivelson_PRB:2008,XuMullerSachdev_PRB:2008,Dai_PNAS:2009,Fernandes_NP:2014,YuSi_PRL:2015,Bohmer-NP2022,Si-Hussey2023}.

The interplay between magnetic and composite orders in the iron-based superconductors has been 
further enriched 
by the 
experimental
findings about unusual
electronic orders.
In optimally hole doped (Ba/Sr,K/Na)Fe$_{2}$As$_{2}$~\cite{Avici_NC:2014,Boehmer_NC:2015,Allred_Osborn_NP:2016,Hassinger_PRB:2016},
a $C_{4}$-symmetric collinear double-Q AFM order has been observed.
Similar magnetic orders preserving the tetragonal symmetry 
have also been
observed in Ni- and Co-doped CaKFe$_{4}$As$_{4}$~\cite{Meier_npjQM:2018},
pressurized FeSe~\cite{Boehmer_PRB:2019}, and heterostructured
Sr$_{2}$VO$_{3}$FeAs~\cite{Ok_Kim_NC:2017}. It is suggested that
the doped CaKFe$_{4}$As$_{4}$ supports a chiral double-Q AFM order~\cite{Meier_npjQM:2018},
and in Sr$_{2}$VO$_{3}$FeAs a hidden order compatible to the collinear
double-Q AFM order exists over a wide temperature regime~\cite{Ok_Kim_NC:2017}.
Theoretical studies~\cite{Lorenzana_PRL:2008,Brydon_PRB:2011,Giovannetti_NC:2011, FernandesKivelsonBerg_PRB:2016,Yu_arXiv:2017}
have considered the collinear and chiral double-Q AFM orders 
through the superposition
of the
$(\pi,0)$ and $(0,\pi)$ AFM orders, 
such that the $C_{4}$ 
symmetry is preserved. 
Intriguingly, these AFM orders
support charge $C_{4}$ \footnote{In this phase the order parameter is the $(\pi,\pi)$ component of the squared spin density. Because it breaks lattice translational symmetry, it should be linearly coupled to a $(\pi,\pi)$ charge order, which is denoted here.} and chiral $C_{4}$ composite
orders, respectively, 
in analogy to the single-Q AFM nucleating the nematic composite order.
The nature of these ordered phases and the transitions among them 
remain to be clarified. This is especially so for the transition
between the single-Q AFM and collinear double-Q AFM orders,
which takes place 
near optimal superconductivity~\cite{Avici_NC:2014,Boehmer_NC:2015}. 
Given that these
two AFM orders 
break different lattice symmetries, the transition is expected to
be first-order. 
However, it 
would be rather difficult to understand the enhanced nematic
fluctuations observed in the collinear double-Q AFM phase~\cite{Frandsen_PRL:2017,Wang_PRB:2018,Song_PRL:2021}
within this na\"{\i}ve picture.

In this paper,
we examine the phase transition between these 
electronic orders
via a Ginzburg-Landau (GL) model. We uncover a closed Lie algebra 
that links
the nematic, charge $C_{4}$, and chiral $C_{4}$ composite order
parameters.
This hidden algebra  allows for a rotation between
the nematic and charge $C_4$ orders 
at the first-order phase transition
that separates the two.
As a result, an emergent $U(1)$ symmetry develops at this first order phase transition, 
and this symmetry
supports
a gapless Goldstone mode. 
As a signature of the emergent symmetry, the susceptibilities of the two corresponding
orders are strongly enhanced close to the transition. We further
show that our results on the emergent 
soft mode 
is robust 
to the presence of the AFM order
and
a variety of perturbations. Finally, we propose
ways to experimentally
detect these effects, analyze the experimental data on the phonon softening in this light,  and discuss the general implications of our results for
the physics of iron-based superconductors and other strongly correlated systems.

\begin{figure}[!t]
\centering \includegraphics[width=0.75\linewidth]{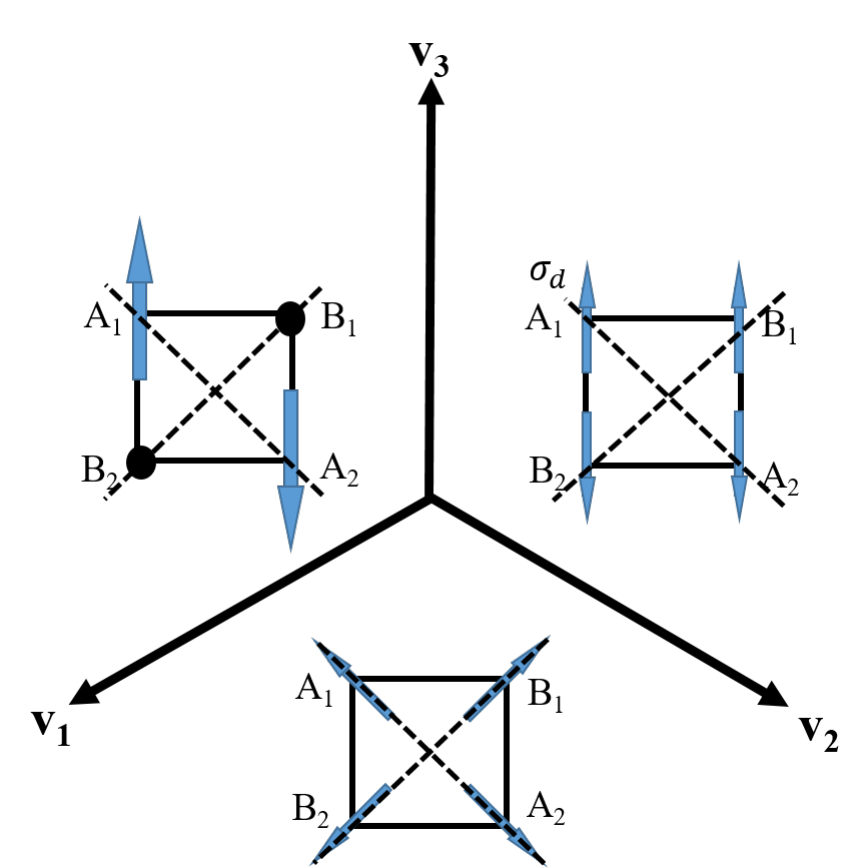}
\caption{\label{fig1}
Ground-state phase diagram of the Ginzburg-Landau model 
studied in this work, viewed 
along the {[}111{]} direction in the $\{v_{1},v_{2},v_{3}\}$
space. The iron square lattice is divided
to four sublatices $A_{1}$, $A_{2}$, $B_{1}$ and $B_{2}$. Sketched
are the spin patterns of the single-Q, the collinear double-Q
and the chiral double-Q AFM states, respectively. Each breaks distinct lattice symmetry and supports a composite order (see text).}
\end{figure}

\noindent
\textit{The $O(N)$
GL model.~} To discuss the various
magnetic and associated composite orders, we adopt the 
description of Ref.~\cite{Yu_arXiv:2017} and construct a
GL
action based on the staggered sublattice magnetic moments $\bm{m}_{A}$
and $\bm{m}_{B}$ on the square lattice $\bm{m}_{A}=\bm{m}_{A_{1}}-\bm{m}_{A_{2}},\bm{m}_{B}=\bm{m}_{B_{1}}-\bm{m}_{B_{2}},$
where the the $A$, $B$ sublattices can be further divided to $A_{1},A_{2}$
and $B_{1},B_{2}$, as illustrated in Fig.~\ref{fig1}. 
The GL action
reads as 
$S=S_{2}+S_{4}$ 
with
\begin{align}
S_{2}= & \frac{1}{2}\sum\limits _{\bm{q},i\omega_{l}}\chi_{0}^{-1}(\bm{q},i\omega_{l})[|\bm{m}_{A}(\bm{q},i\omega_{l})|^{2}+|\bm{m}_{B}(\bm{q},i\omega_{l})|^{2}]\nonumber \\
 & +2w(q_{x}^{2}-q_{y}^{2})\bm{m}_{A}(\bm{q},i\omega_{l})\cdot\bm{m}_{B}(-\bm{q},-i\omega_{l}),\label{magmodel2} \\
S_{4}= & \int_{0}^{\beta}d\tau\int d^{2}x\{v_{1}\left(\bm{m}_{A}\cdot\bm{m}_{B}\right)^{2}+v_{2}\left(\frac{\bm{m}_{A}^{2}-\bm{m}_{B}^{2}}{2}\right)^{2}\nonumber \\
 & +v_{3}\sum\limits _{\alpha<\beta}(m_{A}^{\alpha}m_{B}^{\beta}-m_{B}^{\alpha}m_{A}^{\beta})^{2}\},\label{magmodel4}
\end{align}
where $\bm{m}_{A/B}$ are the staggered magnetic moments on $A/B$ sublattices with generalized $N$ components; $\omega_{l}$ is the bosonic Matsubara
frequency;  $\chi_{0}^{-1}\equiv r+cq^{2}+\gamma|\omega_{l}|$ is
the inverse magnetic susceptibility; $r$ is the mass term; $\gamma$
is the Landau damping rate from coupling to itinerant electrons and $c$ is the squared spin-wave velocity.
In the presence of strong Landau damping, the dynamical exponent $z=2$,
which leads to the effective dimension $d_{eff}=d+z=4$. $w$ controls
the spin anisotropy in the momentum space. 
Here we have generalized the physical $O(3)$
moments $\bm{m}_{A/B}$ to $O(N)$ vectors, and the vector product $\bm{m}_{A}\times\bm{m}_{B}$ to $m_{A}^{\alpha}m_{B}^{\beta}-m_{B}^{\alpha}m_{A}^{\beta}$.
This effective model exhibits $O(N)\times\mathbb{Z}_{2}^{\mathrm{sub}}\times\mathbb{Z}_{2}^{\mathrm{mir}}$
symmetry:  
$O(N)$ refers to spin rotation; 
$\mathbb{Z}_{2}^{\mathrm{sub}}$ 
corresponds to interchanging $A$ and $B$ sublattices $\bm{m}_{A}\leftrightarrow\bm{m}_{B}$; 
and $\mathbb{Z}_{2}^{\mathrm{mir}}$ is generated by the the mirror 
reflection $\sigma_d$ about the diagonal direction of the square lattice (see Fig.~\ref{fig1}).

Compared to its conventional form~\cite{JWu:2016}, the $S_{4}$
term in Eq.~\eqref{magmodel4} has been rewritten 
by three
ordered channels (see below) with quartic couplings $v_{i}$ $(i=1,2,3)$.
This model consists of three magnetic phases:
a single-Q AFM for $v_{1}<v_{2},v_{3}$, a collinear double-Q AFM
for $v_{2}<v_{1},v_{3}$, and a chiral double-Q AFM for $v_{3}<v_{1},v_{2}$,
as
illustrated in Fig.~\ref{fig1}. These
AFM orders support the following
composite orders that break the $\mathbb{Z}_{2}^{\mathrm{sub}}\times\mathbb{Z}_{2}^{\mathrm{mir}}$
lattice symmetries: the nematic order
breaking $\mathbb{Z}_{2}^{\mathrm{mir}}$,
the charge $C_{4}$ order
breaking $\mathbb{Z}_{2}^{\mathrm{sub}}$,
and the chiral $C_{4}$ order
breaking both $\mathbb{Z}_{2}^{\mathrm{sub}}$
and $\mathbb{Z}_{2}^{\mathrm{mir}}$, but preserving their product. These
composite
orders are characterized by the order parameters $\hat{\rho}_1\sim\bm{m}_{A}\cdot\bm{m}_{B}$,
$\hat{\rho}_2\sim(\bm{m}_{A}^{2}-\bm{m}_{B}^{2})/2$, and $\hat{\rho}^{\alpha\beta}_3\sim m_{A}^{\alpha}m_{B}^{\beta}-m_{B}^{\alpha}m_{A}^{\beta}$
$(1\le\alpha<\beta\le N)$, respectively, and transform under different irreducible representations of the $\mathbb{Z}_{2}^{\mathrm{sub}}\times\mathbb{Z}_{2}^{\mathrm{mir}}$ group.

\begin{figure}[!t]
\centering
\subfloat[\label{fig:mexican}]{%
\includegraphics[width=\columnwidth]{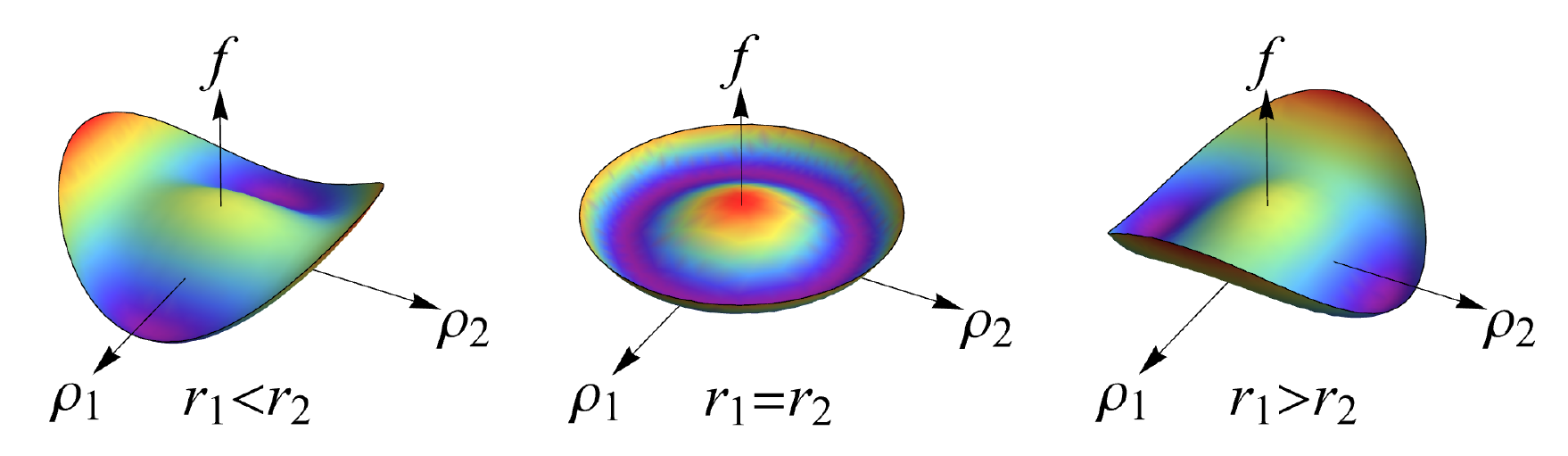}%
}
\hfill
\subfloat[\label{fig:rho}]{%
\includegraphics[width=0.5\columnwidth]{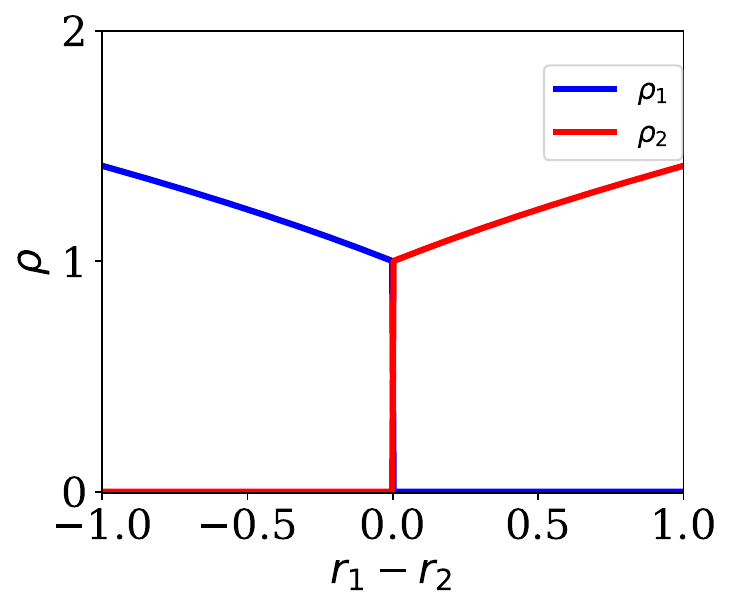}%
}
\hfill
\subfloat[\label{fig:chi}]{%
\includegraphics[width=0.5\columnwidth]{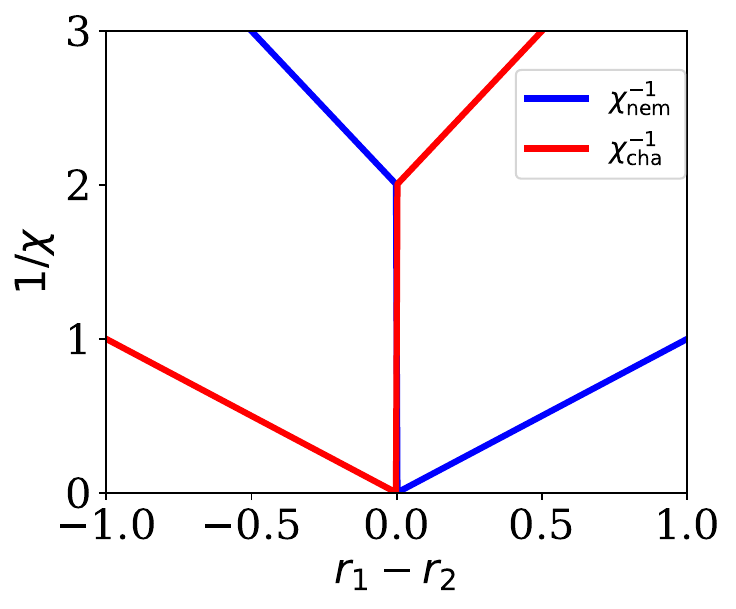}%
}
	\caption{(a) Sketech of the free energy $f$ as a function of
		nematic ($\rho_1$) and charge $C_4$ ($\rho_2$) order parameters, illustrating the Goldstone mode associated with the breaking of the emergent $U(1)$ symmetry. (b) Nematic (blue) and charge $C_{4}$ (red) order parameters and  as a function of
		$r_{1}-r_{2}$. The transition takes place at $r_1=r_2$. (c) Inverse of the nematic (blue) and charge $C_4$ (red)
		susceptibilities as functions of $r_{1}-r_{2}$ in the limit of $q=0$ and $\omega_{n}\rightarrow0$.
	The	$r_{1}<r_{2}$ and $r_{1}>r_{2}$ regimes correspond to nematic and charge
		$C_{4}$ phases, respectively. 
These plots are obtained under the assumption of zero spin anisotropy, which is treated as weak in the main text.}
\label{fig2}
\end{figure}

\noindent
\textit{Emergent $U(1)$ symmetry.}
We first focus on the paramagnetic
phase.
The composite orders emerge
from magnetic fluctuations in the different symmetry channels.
By decomposing the quartic part of the action in terms of Hubbard-Stratonovich
fields $\hat{\rho}_{i}$ $(i=1,2,3)$ of these channels, and integrating out all the
fields of magnetic orders, we obtain an effective Landau theory for the composite
orders (see Supplemental Materials (SM)~\cite{SM}). At the saddle-point level, the expectation values of the Hubbard-Stratonovich fields
$\rho_{i}=\langle \hat{\rho}_{i} \rangle$,
respectively correspond to the three 
composite order parameters.
For anisotropy $w=0$ (the effect of $w$ will be discussed later), the derived
Landau free energy density takes the form 

\begin{align}
f=\sum_{i}r_{i}\rho_{i}^{2}+g(\sum_{i=1}^{3}\rho_{i}^{2})^{2} +g^{\prime}(\rho_{1}^{2}+\rho_{2}^{2})^{2},\label{landau3}
\end{align}
where $\rho_{3}^{2}=\sum_{\alpha<\beta}\rho_{3}^{\alpha\beta}\rho_{3}^{\alpha\beta}$,
and $r_{i}\equiv\frac{1}{2(v_{0}-v_{i})}-\pi(m_{0},T)$ are the
static inverse susceptibilities of the composite order parameters
$\rho_{i}$. Here $v_{0}\equiv\max\{v_{1},v_{2},v_{3}\}$, $m_{0}$ denotes the magnetic mass without composite orders,
and $\pi(m_{0},T),m_{0},g,g^{\prime}$ are functions of $\{r,v_{1},v_{2},v_{3}\}$
(see SM~\cite{SM}). Note that the quadratic coefficient $r_{i}$ has an one-to-one correspondence with $v_{i}$ in Eq.(\ref{magmodel4}), thus when $v_{1}=v_{2}$, $r_{1}=r_{2}$.  It is found that $g^{\prime}=0$ for $N=2$, and $g^{\prime}=g/2$
for $N=3$.
By solving the saddle-point equations~\cite{SM}, we find the transition between any two composite orders to be first-order (see, e.g., Fig.~\ref{fig2}(b)). This is expected within the Landau theory of phase transition, given that the three composite orders break distinct symmetries.
However, quite surprisingly, we find an emergent $U(1)$ symmetry at the transition between the
nematic and charge $C_{4}$ orders: At the transition point $r_{1}=r_{2}$, the free energy density in Eq.~\eqref{landau3} depends only on the total amplitude $\rho=\sqrt{\rho_{1}^{2}+\rho_{2}^{2}}$, and
not on the relative phase angle between $\rho_1$ and $\rho_2$.
Moreover,
we find that for $N=2$, an emergent $U(1)$ symmetry appears at the transition
between any two composite orders.

The emergent $U(1)$ symmetry is not accidental. It 
arises from the $\mathbb{Z}_{2}^{\mathrm{sub}}\times\mathbb{Z}_{2}^{\mathrm{mir}}$
lattice symmetry since the composite order parameters follow a hidden Lie algebra.
  To
see this, we define a $2N$-component vector $\bm{m}=(\bm{m}_{A},\bm{m}_{B})^{\mathrm{T}}$
and rewrite the composite orders in terms of the bilinears of $\bm{m}$.
For the nematic and charge orders, 
$\hat{\rho}_1\sim \bm{m}_{A}\cdot\bm{m}_{B}=\bm{m}^{\mathrm{T}}\bm{N}_{1}\bm{m}$,
$\hat{\rho}_2\sim (\bm{m}_{A}^{2}-\bm{m}_{B}^{2})/2=\bm{m}^{\mathrm{T}}\bm{N}_{2}\bm{m}$.
Here
$\bm{N}_{1}=\tau_{x}\otimes\mathcal{I}_{N}/2$ and $\bm{N}_{2}=\tau_{z}\otimes\mathcal{I}_{N}/2$,
where $\mathcal{I}_{N}$ is the $N\times N$ identity matrix in the
spin space, and $\tau_{i}(i=x,y,z)$ are the Pauli matrices of the
isospin spanned in the sublattice space. 
Defining an
operator $\bm{L}_{3}=-\tau_{y}\otimes\mathcal{I}_{N}/2$, 
one can show that $\bm{N}_{1},\bm{N}_{2}$ and $\bm{L}_{3}$ form an $SO(3)$
Lie algebra, 
 and hence $(\bm{N}_{1},\bm{N}_{2})$ transform as a 2-component vector under
$\bm{L}_{3}$. (For the planar case $N=2$, the hidden Lie algebra is $SO(4)$ as illustrated in section D in SM~\cite{SM}.)
The hidden  
Lie algebra allows continuous rotation between different ordering channels and,
thus, the continuous symmetry. 
To see this, we write the Landau free energy for Eqs.~\eqref{magmodel2} and \eqref{magmodel4} as
\begin{align}
f&=r(\bm{m}^{T}\bm{m})+\frac{v_{3}}{4}(\bm{m}^{T}\bm{m})^{2}\nonumber\\
&-(v_{3}-v_{1})(\bm{m}^{T}\bm{N}_{1}\bm{m})^{2}-(v_{3}-v_{2})(\bm{m}^{T}\bm{N}_{2}\bm{m})^{2},\label{feLandau}
\end{align} 
by
focusing on the spatial and temporal homogeneous case and  
recognizing that $\frac{1}{4}(\bm{m}_{A}^{2}+\bm{m}_{B}^{2})^{2}=\frac{1}{4}(\bm{m}_{A}^{2}-\bm{m}_{B}^{2})^{2}+(\bm{m}_{A}\cdot\bm{m}_{B})^{2}+\sum\limits _{\alpha<\beta}(m_{A}^{\alpha}m_{B}^{\beta}-m_{B}^{\alpha}m_{A}^{\beta})^{2}$. In Eq.~\eqref{feLandau}
the free energy generally has a $\mathbb{Z}_{2}^{\mathrm{sub}}\times\mathbb{Z}_{2}^{\mathrm{mir}}$ symmetry. At $v_{1}=v_{2}$, an additional permutation symmetry with $1\leftrightarrow 2$ emerges if $\bm{N}_1$ and $\bm{N}_2$ are not related, analogous to the case of competing orders between $B_{1g}$, $B_{2g}$ and $A_{2g}$ for incommensurate wave vectors, where only permutation symmetries exist \cite{Wang_PRB:2019}. But when $\bm{N}_1$ and $\bm{N}_2$ are connected by the aforementioned Lie algebra, 
the 
free energy at this point
is invariant under an $O(2)$ rotations in the $2-$dimensional $(\bm{m}_{A},\bm{m}_{B})$
by $\bm{L}_3$. We further note that this 

enhanced continuous symmetry $O(2)\sim U(1)$ is also exact in the
effective theory of Eq.~\eqref{landau3} for composite orders because $v_{1}=v_{2}$ gives $r_{1}=r_{2}$.

We conclude this part by noting on 
an intuitive understanding of the emergent symmetry. We can compare 
the effective model for the composite orders, 
Eq.~(\ref{landau3},\ref{feLandau}), with the anisotropic $XZ$ model,
where a continuous U(1) symmetry emerges at a first-order spin-flop transition~\cite{Beneke2021}. 
The interplay between the nematic order (associated with $\bm{N}_{1}$) and charge $C
_{4}$ order (associated with $\bm{N}_{2}$) 
in Eq.~\eqref{feLandau}
parallels the interplay between the spin components $S_{x}$ and $S_{z}$ in the anisotropic XZ model (see section B in SM~\cite{SM}).

\noindent
\textit{Emergent Goldstone mode at the transition point.} At the nematic to charge $C_4$ transition,
the amplitude $\rho$
is finite (Fig.~\ref{fig2}(b)). This indicates
that the $U(1)$ symmetry is spontaneously broken
(which can only take place at $T=0$ in two dimension but operate at nonzero temperatures in the presence of an interlayer coupling). Associated with this broken symmetry, a Goldstone mode 
emerges.
On each side of the transition, one of
the nematic and charge fluctuations is 
largely enhanced and becomes singular 
at the transition point.
This is shown in the calculated nematic/charge susceptibility:
$\chi_{nem/cha}(x) \sim  \langle\rho_{1/2}(x)\rho_{1/2}(0)\rangle-\langle\rho_{1/2}\rangle^{2}$,
where $x\equiv(\bm{x},t)$.
To obtain the dynamical susceptibilities inside the ordered
phases, we consider fluctuations of the order parameters beyond the saddle-point level, and obtain~\cite{SM}:
\begin{align}\label{nem_cha_sus}
\chi_{nem/cha}(\bm{q},\omega) & =\frac{1}{\delta_{nem/cha}+K\bm{q}^{2}-K_{\tau}(\omega+i\eta)^{2}},
\end{align}
where the coefficients $K$ and $K_{\tau}$ are integrals over the momentum space (section E in SM~\cite{SM}),    $\delta_{nem/cha}=r_{1/2}+12g\overline{\rho}_{1/2}^{2}+4g\overline{\rho}_{2/1}^{2}$
is the mass of nematic/charge fluctuations, and $\overline{\rho}_{1/2}$
is the saddle-point value of $\rho_{1/2}$.
More explicitly, in the disordered phase $(r_{1},r_{2}\geq0)$, $\delta_{nem}=r_{1},\delta_{cha}=r_{2}$;
in the nematic phase $(r_{1}<0,r_{1}<r_{2})$, $\delta_{nem}=-2r_{1}$,
$\delta_{cha}=r_{2}-r_{1}$; and in the charge $C_{4}$ phase $(r_{2}<0,r_{2}<r_{1})$,
$\delta_{nem}=r_{1}-r_{2}$, $\delta_{cha}=-2r_{2}$. Therefore, on  approaching 
the transition in the nematic/charge $C_4$ phase ($r_{1}\rightarrow r_{2}$), $\delta_{\rm{cha/nem}}\rightarrow 0^{+}$, namely,
the charge/nematic
fluctuations become gapless. Remarkably, despite the $z=2$ dynamics of magnetic excitations in the original model of Eqs.~\eqref{magmodel2} and \eqref{magmodel4},
the nematic/charge susceptibility exhibits
distinct dynamics, with $z=1$, as shown in Eq.~\eqref{nem_cha_sus}. 
Across the transition,
the gapless excitation acquires a finite mass, and the susceptibility
exhibits a 
nonzero jump (Fig.\ref{fig2} (c)), consistent with the first-order nature of the transition.

\noindent
\textit{Stability of the emergent symmetry.} The stability of the emergent $U(1)$ symmetry persists in magnetic phases where the spin rotational symmetry is broken; without spin-orbit coupling (SOC), spin and lattice sectors are decoupled, so breaking spin rotational symmetry does not impact the $\mathbb{Z}_{2}^{\mathrm{sub}}\times\mathbb{Z}_{2}^{\mathrm{mir}} \rightarrow U(1)$ enhancement in the lattice sector. This result, verified through large-$N$ analysis of the transition between single-Q and double-Q AFM phases, indicates spontaneous $U(1)$ symmetry breaking with a Goldstone mode, similar to the paramagnetic case (see section F in SM~\cite{SM}).  
In the presence of SOC,
the $U(1)$ symmetry still emerges when magnetic moments align parallel or perpendicular to the crystalline $c$-axis, and is approximately satisfied in other configurations given that,
 in iron-based systems, the
 SOC
 is weaker than magnetic interactions. There are other perturbations that make the $U(1)$ symmetry approximate. For example,
the experimentally observed anisotropic spin fluctuations 
can be characterized by the $w$ term of Eq.~\eqref{magmodel2}, and the $U(1)$ symmetry is only approximate for $w\neq0$. This would introduce a gap in the Goldstone mode, but this gap remains minimal 
(with $(w/c)^4 \approx 10^{-2}$ for iron pnictides), 
and an approximate symmetry is maintained (see section G in SM~\cite{SM}). The eighth order term in the Landau free energy, $(\bm{m}_{A}^{2}-\bm{m}_{B}^{2})^{2}(\bm{m}_{A}\cdot\bm{m}_{B})^{2}$, also breaks
the symmetry, but this term 
is 
negligible near the magnetic transition (see section C in SM~\cite{SM}).

\noindent
\textit{Signatures of the
emergent symmetry.} 
As shown above, a soft mode near the transition survives under the various perturbations in the iron pnictides. 
Approaching the transition from the nematic phase, the
charge $C_4$ excitation becomes soft, and in the charge $C_4$ phase, the nematic excitation is soft. Approaching
the transition point, the soft modes lead to near divergence of the corresponding correlation lengths, and when they meet at the transition, 
a pseudo-Goldstone mode develops.
This divergence in the correlation length can be seen through thermodynamic means, in the form of a divergent Gr\"uneisen ratio (the ratio of the thermal expansion to specific heat), which follows from the general analysis for quantum critical systems~\cite{zhu2003} and parallels the 
consideration for the $XZ$ spin system~\cite{Beneke2021}. We describe this thermodynamic signature in section H in SM\cite{SM}.

Next we 
focus on 
the softening of the symmetry-related in-plane transverse acoustic
(IPTA) phonons~\cite{Li_PRX:2018} across the charge-$C_4$-to-nematic transition. The coupling to the 
nematic fluctuations softens the phonon mode
near the transition, 
\begin{align}
    E=\sqrt{\frac{r_{1c}}{2}\left\lbrace  f(\delta,\tilde{k},\epsilon)-\sqrt{f(\delta,\tilde{k},-\epsilon)^{2}+4\epsilon \tilde{k}^{2}}\right\rbrace} \label{pe}
\end{align}
with
$f(\delta,\tilde{k},\epsilon)=1+\delta+(1+\epsilon)\tilde{k}^{2}$, $\delta=(r_{1}-r_{1c}+12g\overline{\rho}_{1}^{2}+4g\overline{\rho}_{2}^{2})/r_{1c}$; and $\epsilon=c_{T}^{2}/K$; $\tilde{k}=k\sqrt{K/r_{1c}}$. 
Here, $c_{T}$ is the bare velocity of the IPTA phonon; 
$\delta$ is the distance to the nematic transition; $r_{1c}$
is the correction to the nematic mass term from a spin-phonon coupling (section I in SM). The renormalized phonon velocity $\tilde{c}_{T}$, defined in the small momentum limit, is 
$\tilde{c}_{T}=c_{T}\sqrt{\delta/(1+\delta)}$. Near the nematic transition, $\delta\propto\xi_{nem}^{-2}$,   $\tilde{c}_{T}\propto \xi_{nem}^{-1}$, where $\xi_{nem}$ is the nematic
correlation length. Therefore,
 the renormalized phonon velocity is inversely proportional to the nematic
correlation length.

\begin{figure}[tb!]
\centering 
\centering
\subfloat[\label{fig3}]{%
\includegraphics[width=.50\columnwidth]{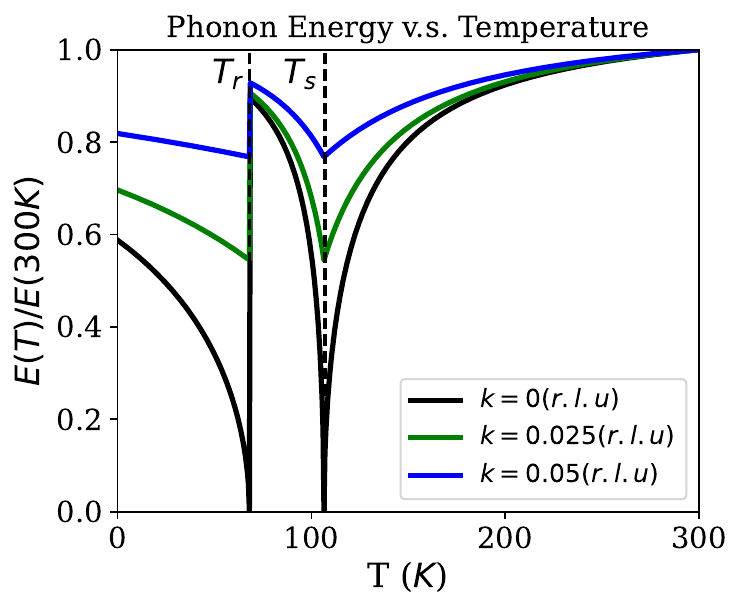}%
}
\hfill
\subfloat[\label{fig4}]{%
\includegraphics[width=0.50\columnwidth]{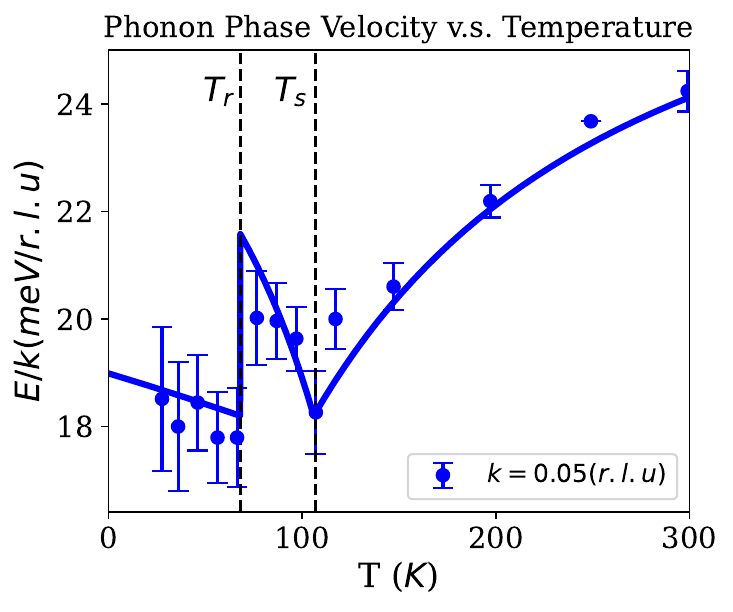}%
}
\caption{
(a) Calculated phonon energy $E(T)/E(300K)$ and (b) measured 
and fitted 
phase velocity $E/k$ of the IPTA phonons 
vs. temperature for Sr$_{0.64}$Na$_{0.36}$Fe$_2$As$_2$ \cite{Song_PRL:2021} (a)
at $k=0$ and two nonzero $k$ values and (b) at $k=0.05 \, r.l.u$ (solid symbols). The fit in (b) is in terms of our analytical results (solid line). The regions $T_{c}<T<T_{r}$, $T_{r}<T<T_{s}$ and $T>T_{s}$ correspond to double-Q AFM, single-Q AFM and tetragonal PM phases, respectively. The theoretical plots are obtained under the assumption of zero spin anisotropy, which is treated as weak in the main text.}
\end{figure}

We compare our theory with a recent
experiment on Sr$_{1-x}$Na$_{x}$Fe$_2$As$_2$, where the softening of the IPTA phonon in the collinear double-Q AFM phase is observed through  inelastic X-ray scattering~\cite{Song_PRL:2021}. 
Our theory is in good agreement with the experimental data, as is illustrated in Fig. \ref{fig4} for $k=0.05(r.l.u)$, where $T_s=107K$ and $T_{s}=68K$ are transition temperatures for nematic and nematic-to-charge-$C_{4}$ transition  respectively.  We also plot the calculated phonon energy using Eq.(\ref{pe}) for $k=0, 0.025,  0.05(r.l.u)$.  Due to limitations imposed by crystal mosaicity and instrument resolution, experimentally accessing smaller values of $k$ is challenging. 
Accordingly,
further experimental studies are needed to better understand the phase boundary across the reentrant charge-$C_4$ phase. We can see that the energy of IPTA phonons is intensively softened across both the nematic transition, as observed in Sr$_{1-x}$Na$_{x}$Fe$_2$As$_2$~\cite{Song_PRL:2021} and also in LiFeAs~\cite{Li_PRX:2018}. A similar softening occurs when approaching the charge-$C_4$ to nematic transition from the charge-$C_4$ side, followed by a sharp jump across the transition point into the nematic phase.
This behavior results from
the diverging nematic correlation length $\xi$, which occurs not only at continuous transitions but also on one side of a first-order transition with enhanced continuous symmetry.
To detect the soft 
charge $C_4$ excitations, one may measure the symmetry-related charge fluctuations
via Raman or RIXS measurements. We leave the details of these and related experimental signatures for future studies.

\noindent
\textit{Discussion and Conclusion.~} The
realization of the single-Q, collinear double-Q, and chiral double-Q AFM phases in LaOFe(As$_{1-x}$,P$_x$) by continuous P doping~\cite{Chmaissem_CP:2022} suggests quasi-degeneracy of the three ground states, and provides the opportunity to systematically examine the related phase transitions, to which our theory is pertinent. As a parallel direction, one can study the $B_{2g}$ nematic order and related composite orders by reformulating the current theory (which is for the $B_{1g}$ nematicity). We expect such a theory to help understand the
$B_{2g}$ nematic fluctuations
that may be relevant to some hole-doped iron pnictides~\cite{Feng:2018, Wang_PRB:2019}.

Our results also shed light on the unconventional superconductivity in iron-based superconductors. 
Recovering the multiorbital itinerant electron degrees of freedom,
our work sets the stage to address how the composite orders affect the superconducting $T_c$ and the pairing structure, which is 
important given that the optimal $T_c$ appears near the single-Q to double-Q AFM transition in some iron pnictides~\cite{Avici_NC:2014, Boehmer_NC:2015}. 

More generally, complex phase competition represents a central theme for a broad array of strongly correlated systems. With suitable generalization, our effective theory
can be applied to other correlated electron systems with four- or six-fold lattice symmetry. 
To illustrate the point with one example, the
proposed triple-Q AFM state in the Kitaev AFM material Na$_2$Co$_2$TeO$_6$ may also contain an interesting structure of composite orders~\cite{ChenLi_PRB:2021}.

In conclusion, we study an effective Ginzburg-Landau model of the 
coupled antiferromagnetic and corresponding composite orders for iron-based superconductors. We show that different composite orders are connected by a hidden Lie algebra, and as a result, a continuous $U(1)$ symmetry emerges at the first-order transition between two ordered phases. Under perturbations where this emergent symmetry becomes approximate, a soft pseudo-Goldstone mode 
still develops.
We have suggested experimental signatures and used our theory to understand the phonon softening observed in the hole-doped iron-based superconductors.
Our
work
sheds new light on the physics of iron-based superconductivity and may lead to new insights into the interplay of quantum phases in a variety of strongly correlated systems.

\acknowledgements We thank P. Dai, W. Ding, Q. Jiang, Y. Li, Z. X. Liu, N. Xi, M. Yi,
and W. Yu for useful discussions. This work has in part been supported
by the National Science Foundation of China Grant number 12174441
and Ministry of Science and Technology of China, National Program
on Key Research Project Grant number 2016YFA0300504 (R.Y., Y.W.).
The work at Rice 
has been supported by
the U.S. Department of Energy, Office of Science, Basic Energy
Sciences, under Award No. DE-SC0018197 (Y.W.).
and the Robert A.\ Welch Foundation
Grant No.\ C-1411 (Q.S.).
Q.S. acknowledges 
the hospitality of the Aspen
Center for Physics,
which is supported by NSF grant No. PHY-2210452.
\bibliographystyle{apsrev4-2}
\bibliography{EMRef}
\clearpage\setcounter{figure}{0} \makeatletter
\global\long\def\thefigure{S\@arabic\c@figure}%
 \onecolumngrid

\section*{ SUPPLEMENTAL MATERIAL -- Emergent symmetry at transition between
intertwined composite orders in iron-based superconductors}

\subsection{Classification of phase transitions in the Ginzburg-Landau theory}\label{Sec:A}

In this section we discuss the classification of phase transitions in the standard Landau theory involving several order parameters by 
studying an effective Ginzburg-Landau model 
with two Ising fields $\rho_{1}$ and $\rho_{2}$.
The Lagrange density of the model reads
\begin{align}
\mathcal{L}= & \frac{1}{2}\left[r_{1}\rho_{1}^{2}+({\partial_{\mu}\rho_{1}})^{2}+r_{2}\rho_{2}^{2}+({\partial_{\mu}\rho_{2}})^{2}\right] 
 +g_{1}\rho_{1}^{4}+g_{2}\rho_{2}^{4}+2g_{12}\rho_{1}^{2}\rho_{2}^{2},\label{Z2}
\end{align}
which has a 
$\mathbb{Z}_{2}\times\mathbb{Z}_{2}$ symmetry. There are two separate transitions corresponding to breaking the two $\mathbb Z_2$ symmetries, respectively. When these two transitions meet, 
the multicritical behavior of this model has been well studied 
\cite{ChaikinLubensky}.
It has been shown that when $g_{1}g_{2}\leq g_{12}^{2}$, there is a bicritical point separating
the disordered phase ($\rho_1=\rho_2=0$) and two ordered phases (with $\rho_1\neq0$, $\rho_2=0$ and vice versa). While the transition
from the disordered phase to 
either of the ordered phases is second-order,
the transition between the two ordered phases along $r_{1}=r_{2}<0$
is essentially first-order. On the other hand, when $g_{1}g_{2}>g_{12}^{2}$,
there is a regime where the two orders 
$\rho_1$ and 
$\rho_2$
coexist. 
The coexisting phase survives up to a tetracritical point. 
Interestingly, when $r_{1}=r_{2}$ and $g_{1}=g_{2}=g_{12}$,
the two Ising fields
combine into a two-component field, and the symmetry
of the system is enhanced from the discrete $\mathbb{Z}_{2}\times\mathbb{Z}_{2}$
to a continuous $U(1)$ symmetry. In this special case, as the continuous
symmetry is spontaneously broken, the longitudinal fluctuation and
the transverse fluctuation around the broken fields lead to two different
excitations: a Higgs mode with a gapped excitation and a Goldstone
mode with a gaplesss excitation. 
However, it usually requires fine-tuning to approach to 
Therefore, the bicritical point with an enhanced symmetry is hardly realized in model systems.
But as shown in the manuscript, 
for the composite orders in iron-based superconductors, an emergent symmetry can arise at the transition point without fine tuning.

\subsection{Anisotropic XZ model}
In the context of our study on iron-based superconductors and the emergent $U(1)$ symmetry, we draw an analogy with the anisotropic XZ model. The Hamiltonian of the anisotropic XZ model is given by:
\begin{equation}
H = -J\sum_{\langle i, j \rangle} \left( (1+\Delta)S_i^x S_j^x + (1-\Delta)S_i^z S_j^z \right)
\end{equation}
where $J>0$ is the coupling constant, $\Delta$ is the measure of the anisotropy in the spin interactions. It differentiates the coupling strength in the $x$-direction from that in the $z$-direction. The symmetry of the Hamiltonian is $\mathbb{Z}_{2}^{x}\times\mathbb{Z}_{2}^{z}$, where the first $\mathbb{Z}_{2}^{x}$ is the symmetric for $S^{x}$, and second $\mathbb{Z}_{2}^{z}$ represents the symmetry for $S^{z}$.

Regarding the phases:
\begin{itemize}
    \item For $\Delta>0$, The coupling in the $x$-direction is stronger. $\mathbb{Z}_{2}^{x}$ symmetry is broken. This can lead to a phase where the spins are ordered that prioritizes the $x$-component of the spins: $\langle S^{x}\rangle \neq 0$. In the context of the main text, this is analogous to nematic ordering.
    \item For $\Delta<0$, The coupling in the $z$-direction is stronger. $\mathbb{Z}_{2}^{z}$ symmetry is broken. Spins are ordered along $z$-component of the spins: $\langle S^{z}\rangle \neq 0$. This is analogous to charge $C_4$ ordering.
    \item For $\Delta=0$, The system is isotropic with equal coupling in both the $x$ and $z$ directions. The symmetry of the Hamiltonian is enhanced to $U(1)$. When $U(1)$ is broken, one direction is picked to be ordered while the spin excitation along the other perpendicular direction is gapless, featuring the emergence of Goldstone mode.
\end{itemize}

\subsection{Stability of the emergent continuous symmetry.}
The $U(1)$ symmetry
is stable 
in the presence of magnetic orders.
In a magnetic phase, besides the $\mathbb{Z}_{2}^{\mathrm{sub}}\times\mathbb{Z}_{2}^{\mathrm{mir}}$ lattice symmetry,
the spin $O(N)$ symmetry is also broken. 
However, without any spin-orbit coupling (SOC), the spin and lattice sectors are decoupled. Therefore, breaking the spin rotational symmetry does not affect the $\mathbb{Z}_{2}^{\mathrm{sub}}\times\mathbb{Z}_{2}^{\mathrm{mir}}\rightarrow U(1)$ symmetry enhancement in the lattice sector. Note that this is inherent in the Lie algebra, where the transition is where the magnetic transition takes place.

We have verified these conclusions from
explicit calculations in
the large-$N$ 
framework for the transition between the single-Q and collinear double-Q AFM phases~\cite{SM}.

By taking into account the fluctuations about the saddle point, we still find the presence of a Goldstone mode in the composite sector. This clearly evidences a spontaneous $U(1)$ symmetry breaking, as in the paramagnetic case. 

A finite 
SOC reduces the continuous $O(N)$ spin rotational symmetry 
to 
discrete space group symmetry. 
We find that the $U(1)$ symmetry still emerges when the magnetic moments in the two ordered phases are both along or both perpendicular to the crystalline $c$-axis. In more general cases, this $U(1)$ symmetry 
is approximately satisfied, given that the SOC is an order of magnitude smaller than the magnetic energy scale.

Higher-order terms in the Ginzburg-Landau model could also potentially break the $U(1)$ symmetry. We find that the sixth-order term $[v_{61}(\bm{m}_{A}\cdot\bm{m}_{B})^{2}+v_{62}(\bm{m}_{A}^{2}-\bm{m}_{B}^{2})^{2}/4](\bm{m}_{A}^{2}+\bm{m}_{B}^{2})$ preserves the emergent symmetry at the transition point (shifted by $v_{61}$ and $v_{62}$), but the eighth-order term $(\bm{m}_{A}^{2}-\bm{m}_{B}^{2})^{2}(\bm{m}_{A}\cdot\bm{m}_{B})^{2}$ breaks it. Nevertheless, this eighth-order term is considered to be negligible near the magnetic transitions $|\bm{m}_{A}|,|\bm{m}_{B}|\ll 1$.

There are other perturbations that make the $U(1)$ symmetry approximate. For example, 
the experimentally observed anisotropic spin fluctuations 
can be characterized by the $w$ term of Eq.~\eqref{magmodel2}, and the $U(1)$ symmetry is only approximate for $w\neq0$. Consequently, the Goldstone mode at the transition is gapped out.

But 
when $w\ll c$, a pseudo-Goldstone mode with
the gap 
of the $(w/c)^{4}$ order remains~\cite{SM}.

For typical 
iron pnictides,
$w/c\approx1/3$~\cite{JWu:2016}, which
leads to the gap ratio of the pseudo-Goldstone to the Higgs mode to be
at the order of $(w/c)^{4}\approx10^{-2}$, which is negligibly small.

\subsection{Hidden Lie algebra}
We have shown in the main text that there is a hidden $SO(3)$ Lie algebra for $N>2$. For completeness, we restate the proof here:
we define a $2N$-component vector $\bm{m}=(\bm{m}_{A},\bm{m}_{B})^{\mathrm{T}}$
and rewrite the composite orders in terms of the bilinears of $\bm{m}$.
For the nematic and charge orders,
$\hat{\rho}_1\sim \bm{m}_{A}\cdot\bm{m}_{B}=\bm{m}^{\mathrm{T}}\bm{N}_{1}\bm{m}$,
$\hat{\rho}_2\sim (\bm{m}_{A}^{2}-\bm{m}_{B}^{2})/2=\bm{m}^{\mathrm{T}}\bm{N}_{2}\bm{m}$.
Here 
$\bm{N}_{1}=\tau_{x}\otimes\mathcal{I}_{N}/2$ and $\bm{N}_{2}=\tau_{z}\otimes\mathcal{I}_{N}/2$, where $\mathcal{I}_{N}$ is the $N\times N$ identity matrix in the
spin space, and $\tau_{i}$ 
 with $i=x,y,z$ are the Pauli matrices of the
isospin spanned in the sublattice space. By defining an
operator $\bm{L}_{3}=-\tau_{y}\otimes\mathcal{I}_{N}/2$. It is easy
to check that $\bm{N}_{1},\bm{N}_{2}$ and $\bm{L}_{3}$ form an $SO(3)$
Lie algebra,
 and hence $(\bm{N}_{1},\bm{N}_{2})$ transform as a 2-component vector under
$\bm{L}_{3}$.   

For $N=2$, the above $SO(3)$ algebra expands to $SO(4)$. In this
case, the chiral order parameter becomes a scalar and can be expressed
as $\bm{m}^{\mathrm{T}}\bm{N}_{3}\bm{m}$, where $\bm{N}_{3}=\tau_{y}\otimes\sigma_{y}/2$,
and $\sigma_{i}(i=x,y,z)$ are the Pauli matrices in the spin space.
Now we define two more operators $\bm{L}_{1}=-\tau_{x}\otimes\sigma_{y}/2$
and $\bm{L}_{2}=-\tau_{z}\otimes\sigma_{y}/2$. One can verify that
$\bm{L}_{i}$ and $\bm{N}_{i}$ span an $SO(4)$ Lie algebra: $[\bm{L}_{i},\bm{L}_{j}]=i\epsilon_{ijk}\bm{L}_{k}$,
$[\bm{L}_{i},\bm{N}_{j}]=i\epsilon_{ijk}\bm{N}_{k}$, and $[\bm{N}_{i},\bm{N}_{j}]=i\epsilon_{ijk}\bm{L}_{k}$.
Note that $\bm{L}_{i}$ are the generators of an $SO(3)$ subgroup,
and $(\bm{N}_{1},\bm{N}_{2},\bm{N}_{3})$ transform as a 3-component
vector under the $SO(3)$ generators $\bm{L}_{i}$. It 
is then 
easy to show that, constrained by the larger algebra, $U(1)$ symmetry emerges at transition between any two composite orders.

\subsection{Effective field theory for the composite orders in the paramagnetic phase
}

Here, we show that in the Ginzburg-Landau model 
describing the magnetic fluctuations 
of iron-based superconductors, the phase transition
between the nematic 
and charge $C_{4}$ composite orders
is first-order. We further calculate the susceptibilities corresponding to the composite order parameters, and show that the system exhibits an emergent Goldstone mode due to breaking the emergent $U(1)$ symmetry at the transition.

The action of the Ginzburg-Landau model
has the form $S=S_{2}+S_{4}$, where
\begin{align}
S_{2}= & \frac{1}{2}\sum\limits _{\bm{q},i\omega_{l}}\chi_{0}^{-1}(\bm{q},i\omega_{l})[|\bm{m}_{A}(\bm{q},i\omega_{l})|^{2}+|\bm{m}_{B}(\bm{q},i\omega_{l})|^{2}]+2w(q_{x}^{2}-q_{y}^{2})\bm{m}_{A}(\bm{q},i\omega_{l})\cdot\bm{m}_{B}(-\bm{q},-i\omega_{l}) \label{SM_S2} \\
S_{4}= & \int_{0}^{\beta}d\tau\int d^{2}x \{u_{1}\left(\bm{m}_{A}^4+\bm{m}_{B}^4\right) +u_{2}\left(\bm{m}_{A}^{2}\bm{m}_{B}^{2}\right)-u_{I}(\bm{m}_{A}\cdot \bm{m}_{B})^{2}\}.\label{SM_S4}
\end{align}
Eq.~\eqref{SM_S4} gives the conventional form of the quartic part $S_4$ of the action~\cite{JWu:2016}. Rewriting $S_4$ in terms of the three composite order parameters, we obtain the form in the main text,
\begin{align}
S_{4}= & \int_{0}^{\beta}d\tau\int d^{2}x\{v_{1}\left(\bm{m}_{A}\cdot\bm{m}_{B}\right)^{2}+v_{2}\left(\frac{\bm{m}_{A}^{2}-\bm{m}_{B}^{2}}{2}\right)^{2}
+v_{3}\sum\limits_{\alpha<\beta}\left(m_{A}^{\alpha}m_{B}^{\beta}-m_{B}^{\alpha}m_{A}^{\beta}\right)^{2}\},
\end{align}
where $v_1$, $v_2$, and $v_3$ are linear combinations of $u_1$, $u_2$, and $u_I$. Here, we have rewritten $\bm{m}_{A}\times\bm{m}_{B}$ into the matrix form $\sum\limits_{\alpha<\beta} \left(m_{A}^{\alpha}m_{B}^{\beta}-m_{B}^{\alpha}m_{A}^{\beta}\right)$, which can be easily generalized to the case when $\bm{m}_{A/B}$ has $N$ components. In the following, $\bm{m}_{A}\times\bm{m}_{B}$ will be used as a convention for this general matrix form.
For a physical system, $S_4$ must be positively definite. This puts a strong constraint to the values of $v_i$. Without losing generality, we can define $v_{0}=\max\{v_{1},v_{2},v_{3}\}+\delta$, where $\delta$ is
an infinitesimally positive number to
regulate the three channels.
Then
$S_{4}$ can be rewritten into
\begin{align}
S_{4}=\int_{0}^{\beta}d\tau d^{2}x\left\lbrace v_{0}\left(\frac{\bm{m}_{A}^{2}+\bm{m}_{B}^{2}}{2}\right) -(v_{0}-v_{1})\left(\bm{m}_{A}\cdot\bm{m}_{B}\right)^{2} -(v_{0}-v_{2})\left(\frac{\bm{m}_{A}^{2}-\bm{m}_{B}^{2}}{2}\right)^{2} -(v_{0}-v_{3})\sum\limits_{\alpha<\beta}\left(m_{A}^{\alpha}m_{B}^{\beta}-m_{B}^{\alpha}m_{A}^{\beta}\right)^{2}\right\rbrace.
\end{align}
We then generalize the spin rotational symmetry to $O(N)$, and perform a standard large-$N$ analysis to obtain the saddle-point solution and fluctuations around the saddle point.

In the large-$N$ approach, we first rescale $\{v_{0},v_{1},v_{2},v_{3}\}$ to $\{v_{0}/N,v_{1}/N,v_{2}/N,v_{3}/N\}$.
After 
applying a Hubbard-Stratonovich transformation, we obtain
\begin{align}
-r\left(\frac{\bm{m}_{A}^{2}+\bm{m}_{B}^{2}}{2}\right)-\frac{v_{0}}{N}\left(\frac{\bm{m}_{A}^{2}+\bm{m}_{B}^{2}}{2}\right)^{2} & \longrightarrow N\frac{(i\lambda-r)^{2}}{4v_{0}}-i\lambda\left(\frac{\bm{m}_{A}^{2}+\bm{m}_{B}^{2}}{2}\right)\\
\frac{v_{0}-v_{1}}{N}(\bm{m}_{A}\cdot\bm{m}_{B})^{2} & \longrightarrow-\frac{N\rho_{1}^{2}}{4(v_{0}-v_{1})}-\rho_{1}\bm{m}_{A}\cdot\bm{m}_{B}\\
\frac{v_{0}-v_{2}}{N}\left(\frac{\bm{m}_{A}^{2}-\bm{m}_{B}^{2}}{2}\right)^{2} & \longrightarrow-\frac{N\rho_{2}^{2}}{4(v_{0}-v_{2})}-\rho_{2}\left(\frac{\bm{m}_{A}^{2}-\bm{m}_{B}^{2}}{2}\right)\\
\frac{v_{0}-v_{3}}{N}\sum\limits_{\alpha<\beta}\left(m_{A}^{\alpha}m_{B}^{\beta}-m_{B}^{\alpha}m_{A}^{\beta}\right)^{2} & \longrightarrow-\frac{N\bm{\rho}_{3}^{2}}{4(v_{0}-v_{3})}-\sum_{\alpha<\beta}\rho_{3}^{\alpha\beta}\left(m_{A}^{\alpha}m_{B}^{\beta}-m_{B}^{\alpha}m_{A}^{\beta}\right)
\end{align}
where $\bm{\rho}_{3}^{2}=\sum_{\alpha<\beta}\rho_{3}^{\alpha\beta}\rho_{3}^{\alpha\beta}$, to the leading order in $1/N$, $i\lambda=r+v_{0}\langle\bm{m}_{A}^{2}+\bm{m}_{B}^{2}\rangle/N$,
refers to the renormalized mass
associated with $\bm{m}_{A}^{2}+\bm{m}_{B}^{2}$. 
$\rho_{1}=-2(v_{0}-v_{1})\langle\bm{m}_{A}\cdot\bm{m}_{B}\rangle/N$,
$\rho_{2}=-(v_{0}-v_{2})\langle\bm{m}_{A}^{2}-\bm{m}_{B}^{2}\rangle/N$
and $\rho_{3}^{\alpha\beta}=-2(v_{0}-v_{3})\langle m_{A}^{\alpha}m_{B}^{\beta}-m_{B}^{\alpha}m_{A}^{\beta}\rangle/N$
are the three Hubbard-Statonovich fields, which respectively refer to the Ising-nematic,
the charge $C_{4}$, and 
the 
chiral $C_{4}$ order parameters.

We first consider the effective model for composite orders in the paramagnetic phase, \textit{i.e.}, without any magnetic order.
In this case, the 
fields $m_{A}$ and $m_{B}$
are pure fluctuations with finite magnetic mass, 
and can be safely integrated
out. 
This leads to the following effective action:
\begin{align}
S_{eff}= & N\int_{0}^{\beta}d\tau\int d^{d}x\left\lbrace -\frac{(i\lambda-r)^{2}}{4v_{0}} +\frac{\rho_{1}^{2}}{4(v_{0}-v_{1})} +\frac{\rho_{2}^{2}}{4(v_{0}-v_{2})} +\frac{\bm{\rho}_{3}^{2}}{4(v_{0}-v_{3})}\right\rbrace +\frac{1}{2}tr\ln{\bm{D}^{-1}},\label{eff}
\end{align}
where
\begin{flalign}
tr\ln{\bm{D}^{-1}}=\ln{det{\begin{pmatrix}\bar{\chi}_{0}^{-1}(x,\tau)+i\lambda+\rho_{2} & \rho_{1}+\bm{\rho}_{3}\\
\rho_{1}-\bm{\rho}_{3} & \bar{\chi}_{0}^{-1}(x,\tau)+i\lambda-\rho_{2}
\end{pmatrix}}}\label{matrix}
\end{flalign}
with $\bar{\chi}_{0}^{-1}=\chi_{0}^{-1}-r$. To the leading order of
$1/N$, the action is 
approximated by a 
saddle point 
with $\{m,\overline{\rho}_{1},\overline{\rho}_{2},\overline{\bm{\rho}}_{3}\}$
, where $m,\bar{\rho_{1}},\bar{\rho}_{2},\bar{\bm{\rho}}_{3}$ respectively
denotes for the saddle-point values of the effective magnetic mass, and order parameters of the nematic, charge, 
and chiral
channels.
In the following, $\rho_i$ refers to the saddle-point value $\bar{\rho}_i$ for simplicity. 
We focus on the physical cases $N=2$ and $N=3$.
For $N=2$, the saddle point equations read 
\begin{align}
\frac{m-r}{v_{0}} & =I(m-\rho,T)+I(m+\rho,T)\label{me}\\
\frac{\rho_{1}}{v_{0}-v_{1}} & =\frac{\rho_{1}}{\rho}(I(m-\rho,T)-I(m+\rho,T))\label{1e}\\
\frac{\rho_{2}}{v_{0}-v_{2}} & =\frac{\rho_{2}}{\rho}(I(m-\rho,T)-I(m+\rho,T))\label{2e}\\
\frac{\rho_{3}}{v_{0}-v_{3}} & =\frac{\rho_{3}}{\rho}(I(m-\rho,T)-I(m+\rho,T)),\label{3e}
\end{align}
where
$\rho=\sqrt{\rho_{1}^{2}+\rho_{2}^{2}+\rho_{3}^{2}}$,
\begin{align}
I(r,t) & =\frac{1}{\beta V}\sum\limits _{\bold{q},i\omega_{l}}\frac{1}{r+cq^{2}+\gamma|\omega_{l}|}\nonumber \\
 & =\frac{1}{4\pi^{2}c\gamma}\left\lbrace (1+\ln(\gamma\Omega))c\Lambda^{2}+r\ln{r}-(r+c\Lambda^{2})\ln(r+c\Lambda^{2})\right\rbrace \left.-\frac{t}{4\pi^{2}c\gamma}\left\lbrace \frac{x}{t}-\frac{x}{t}\ln\frac{x}{t}+\frac{1}{2}\ln{\frac{x}{2\pi t}+\ln{\Gamma\left(\frac{x}{t}\right)}}\right\rbrace \right|_{x=r}^{x=r+c\Lambda^{2}}.
\end{align}
with $t=2\pi\gamma T$ is 
the reduced temperature, 
$\Omega$ and $\Lambda$
are frequency and momentum cutoffs, respectively, and $\Gamma(z)$ is
the Gamma function. 
The saddle-point equations for $N=3$ has a similar form, and the only difference is that $\bm{\rho}_{3}$ 
is identical to a scalar for $N=2$ 
but a vector for $N=3$.

By solving the saddle-point equations \eqref{me}-\eqref{3e}, one can write $m$ in terms of $\{r,v_{0},\rho\}$, and when further assuming
$\rho\ll m$, 
the free energy density $f=S_{eff}^{(0)}/N\beta V$ 
can then be expressed in terms of the composite order parameters
\begin{align}
f
\approx f_{0}+\sum_{i}r_{i}\rho_{i}^{2}+g(\sum_{i=1}^{3}\rho_{i}^{2})^{2}+g^{\prime}(\rho_{1}^{2}+\rho_{2}^{2})^{2}.\label{fden}
\end{align}
where $f_{0}$
is a constant term, $g^\prime=0$ for $N=2$ and $g^\prime=g/2$ for $N=3$.
Either minimizing the free energy in Eq.~\eqref{fden} or solving the saddle-point equations \eqref{me}-\eqref{3e}, it is easy to show that for general $\{v_i\}$ values, there can be two types of solutions: a disordered phase corresponding to all $\rho_i=0$, and an ordered phase corresponding to only one $\rho_i\neq0$. Note that according to the criterion in Sec.~\ref{Sec:A}, there is no coexisting regime between two ordered phases unless at the transition point. It has been shown in the main text that the system exhibits an emergent $U(1)$ symmetry right at the transition point ($r_1=r_2$) between the $\rho_1$ (nematic) and $\rho_2$ (charge $C_4$) ordered phases. This is obviously shown in the free energy of Eq.~\eqref{fden}, which depends only on the total amplitude $\rho$. Minimizing the free energy one finds that $\rho\neq0$ at the transition point. Actually $\rho$ changes continuously across the transition. However, the transition is first-order because the two order parameters $\rho_1$ and $\rho_2$ jump abruptly between zero and $\rho$ at this point.

We then calculate the susceptibility associated to each order parameter. Without loss of generality, we focus on $N=2$ case. We assume $\rho$ is nonzero but small ($\rho\ll m$ near the transition), then the spacial and temperal part of the field can be
determined by expanding Eqn.(\ref{eff}). We obtain
\begin{align}
S_{eff}= & \frac{1}{2}\sum_{\bm{q},i\omega_{l}}\chi_{nem}^{-1}(\bm{q},i\omega_{l})|\rho_{1}(\bm{q},i\omega_{l})|^{2}+\chi_{cha}^{-1}(\bm{q},i\omega_{l})|\rho_{2}(\bm{q},i\omega_{l})|^{2}+\chi_{chi}^{-1}(\bm{q},i\omega_{l})|\rho_{3}(\bm{q},i\omega_{l})|^{2}\nonumber \\
 & +g\int d\tau d^{2}x (\sum_{i=1}^{3}\rho_{i}^{2})^{2},
\label{field theory2}\end{align}
where the susceptibilities of the three channels take the following form:
\begin{align}
\chi_{nem}^{-1}(\bm{q},i\omega_{l})=\frac{1}{2(v_{0}-v_{1})}-\pi(\bm{q},i\omega_{l}),\\
\chi_{cha}^{-1}(\bm{q},i\omega_{l})=\frac{1}{2(v_{0}-v_{2})}-\pi(\bm{q},i\omega_{l}),\\
\chi_{chi}^{-1}(\bm{q},i\omega_{l})=\frac{1}{2(v_{0}-v_{3})}-\pi(\bm{q},i\omega_{l}).
\end{align}
Here
\begin{align}
\pi(\bm{q},i\omega_{l})=\frac{1}{(2\pi)^{3}}\int_{0}^{\Lambda}d^{2}k & \int_{-\infty}^{\infty}d\omega_{n}\frac{1}{m+c(\bm{k}+\bm{q})^{2}+|\omega_{n}+\omega_{l}|}\frac{1}{m+c\bm{k}^{2}+|\omega_{n}|}\nonumber \\
=\frac{1}{(2\pi)^{3}}\int_{0}^{\Lambda}d^{2}k & \frac{\ln({m+c(\bm{k}+\bm{q})^{2}+\omega_{l}})-\ln({m+c\bm{k}^{2}})}{\omega_{l}+c(\bm{k}+\bm{q})^{2}-c\bm{k}^{2}}+\frac{\ln({m+c(\bm{k}+\bm{q})^{2}})-\ln({m+c\bm{k}^{2}}+\omega_{l})}{-\omega_{l}+c(\bm{k}+\bm{q})^{2}-c\bm{k}^{2}}\nonumber \\
 & +\frac{\ln(1+\frac{\omega_{l}}{{m+c(\bm{k}+\bm{q})^{2}}})+\ln(1+\frac{\omega_{l}}{{m+c\bm{k}^{2}}})}{2m+\omega_{l}+c(\bm{k}+\bm{q})^{2}+c\bm{k}^{2}}\nonumber \\
=\frac{2}{(2\pi)^{3}}\int_{0}^{\Lambda}d^{2}k & \frac{1}{m+ck^{2}}-\frac{c}{(2\pi)^{3}}\int_{0}^{\Lambda}d^{2}k\frac{m+3ck^{2}}{(m+ck^{2})^{3}}q^{2}-\frac{1}{3(2\pi)^{3}}\int_{0}^{\Lambda}d^{2}k\frac{1}{(m+ck^{2})^{3}}\omega_{l}^{2}
\end{align}

Denoting
\begin{align}
	r_{i}=&\frac{1}{2(v_{0}-v_{i})}-\frac{2}{(2\pi)^{3}}\int_{0}^{\Lambda}d^{2}k \frac{1}{m+ck^{2}},\\
	K_{x}=&\frac{c}{(2\pi)^{3}}\int_{0}^{\Lambda}d^{2}k\frac{m+3ck^{2}}{(m+ck^{2})^{3}},\\K_{\tau}=&\frac{1}{3(2\pi)^{3}}\int_{0}^{\Lambda}d^{2}k\frac{1}{(m+ck^{2})^{3}}.,\\g=&\frac{v_{0}(\frac{\partial\pi(0,0)}{\partial m})^{2}}{1+2v_{0}\pi(0,0)}-\frac{1}{6}\frac{\partial^{2}\pi(0,0)}{\partial m^{2}},
\end{align}
we arrive to the expression in Eq.~(5) of the main text.

\subsection{Effective field theory for composite orders in the presence of magnetic orders}

In this section we show that our results on the emergent symmetry and Goldstone
mode at the transition between two composite orders still hold even in the presence of magnetic orders.
We consider the transition between the $C_{2}$ single-Q and the $C_{4}$ collinear double-Q AFM phases. 
For this purpose it is convenient to choose $v_{0}\equiv\max\{v_{1},v_{2},v_{3}\}=v_{3}$. Note that the two magnetic orders break distinct lattice symmetries, so the transition between the two magnetic orders and the one between composite orders are the same one.

We perform a large-$N$ analysis for this transition. First set $\bm{m}_{A/B}=(\sqrt{N}\sigma_{A/B},\bm{\pi}_{A/B}^{T})^{T}$,
where $\sigma_{A/B}$ denotes for the longitudial static 
part of the sublatice magnetizations, 
and $\bm{\pi}_{A/B}$ refer to the transverse fluctuations 
containing $N-1$ components. Following similar procedure as in the paramagnetic case, we can separate the action into two parts, consisting of either $\sigma_{A/B}$ or $\bm{\pi}_{A/B}$.

\begin{align}
S= & N\int_{0}^{\beta}d\tau d^{d}x\left\lbrace -\frac{(i\lambda-r)^{2}}{4v_{3}}+\frac{\rho_{1}^{2}}{4(v_{3}-v_{1})}+\frac{\rho_{2}^{2}}{4(v_{3}-v_{2})}+\lambda\left(\frac{\sigma_{A}^{2}+\sigma_{B}^{2}}{2}\right)+\rho_{1}\sigma_{A}\sigma_{B}+\rho_{2}\left(\frac{\sigma_{A}^{2}-\sigma_{B}^{2}}{2}\right)\right\rbrace +S_{\pi_{A},\pi_{B}}
\end{align}
with
\begin{flalign}
S_{\pi_{A},\pi_{B}} & =\int_{0}^{\beta}d\tau\int d^{d}x\left\lbrace \frac{1}{2}\bm{\pi}_{A}^{T}(\bar{\chi}_{0}^{-1}+\lambda+\rho_{2})\bm{\pi}_{A}+\frac{1}{2}\bm{\pi}_{B}^{T}(\bar{\chi}_{0}^{-1}+\lambda-\rho_{2})\bm{\pi}_{B}+\rho_{1}\bm{\pi}_{A}^{T}\bm{\pi}_{B}\right\rbrace \nonumber \\
 & =\frac{1}{2}\begin{pmatrix}\bm{\pi}_{A}^{T} & \bm{\pi}_{B}^{T}\end{pmatrix}\begin{pmatrix}\bar{\chi}_{0}^{-1}+i\lambda+\rho_{2} & \rho_{1}\\
\rho_{1} & \bar{\chi}_{0}^{-1}+i\lambda-\rho_{2}
\end{pmatrix}\begin{pmatrix}\bm{\pi}_{A}\\
\bm{\pi}_{B}
\end{pmatrix}
\end{flalign}
Here $\rho_i$ are the fields for the composite orders as defined before. We then 
integrate out the fluctuating fields $\bm{\pi}_{A}$ and $\bm{\pi}_{B}$, 
and get an effective action including $\rho_i$ and $\sigma_{A/B}$ only,
\begin{align}
S_{eff}= & N\int_{0}^{\beta}d\tau\int d^{d}x\left\lbrace -\frac{(i\lambda-r)^{2}}{4v_{3}}+\frac{\rho_{1}^{2}}{4(v_{3}-v_{1})}+\frac{\rho_{2}^{2}}{4(v_{3}-v_{2})}+i\lambda\left(\frac{\sigma_{A}^{2}+\sigma_{B}^{2}}{2}\right)+\rho_{1}\sigma_{A}\sigma_{B}+\rho_{2}\left(\frac{\sigma_{A}^{2}-\sigma_{B}^{2}}{2}\right)\right\rbrace \nonumber \\
 & +\frac{N-1}{2}tr\ln((\bar{\chi}_{0}^{-1}+i\lambda)^{2}-\rho_{1}^{2}-\rho_{2}^{2}).
\end{align}

The free energy can then be derived, which leads to the following 
saddle-point equations, 
\begin{align}
\frac{\sigma_{A}^{2}+\sigma_{B}^{2}}{2}-\frac{m-r}{2v_{3}}+\frac{1}{2}(I(m-\rho)+I(m+\rho))=0\label{mag1}\\
\sigma_{A}\sigma_{B}+\frac{\rho_{1}}{2(v_{3}-v_{1})}-\frac{\rho_{1}}{2\rho}(I(m-\rho)-I(m+\rho))=0\label{mag2}\\
\frac{\sigma_{A}^{2}-\sigma_{B}^{2}}{2}+\frac{\rho_{2}}{2(v_{3}-v_{2})}-\frac{\rho_{2}}{2\rho}(I(m-\rho)-I(m+\rho))=0\label{mag3}\\
\rho_{1}\sigma_{B}+(m+\rho_{2})\sigma_{A}=0\label{mag4}\\
\rho_{1}\sigma_{A}+(m-\rho_{2})\sigma_{B}=0.\label{mag5}
\end{align}
Here $i\lambda$ is replaced by the effective mass $m$ at the saddle-point level.
From Eqs.{(\ref{mag4},\ref{mag5})}, we get $\sigma_{A}\sigma_{B}=-\sigma^{2}\rho_{1}/\rho$
and $(\sigma_{A}^{2}-\sigma_{B}^{2})/2=-\sigma^{2}\rho_{2}/\rho$,
where $\rho=\sqrt{\rho_{1}^{2}+\rho_{2}^{2}}$, $\sigma^{2}=(\sigma_{A}^{2}+\sigma_{B}^{2})/2$.
Moreover, in order to have a nonzero solution for $\sigma_{A}$ and $\sigma_{B}$,
we should have $m=\rho$. Substituting these expressions back to Eqs.{(\ref{mag1},\ref{mag2},\ref{mag3})},
we find the nematic order parameter and the charge order parameter
are dual to each other, and the transition between them is first-order, same as in the paramagnetic case.

The above saddle-point equations can be solved analytically. In the collinear
$C_{2}$ AFM phase,
\begin{align}
\rho_{1} & =\pm a_{1}e^{W_{0}\left(\frac{r_{c}-r}{4\overline{v}_{3}a_{1}}\right)}\stackrel{r\rightarrow r_{c}^{-}}{\longrightarrow}\pm a_{1},\rho_{2}=0,\\
\sigma_{A} & =-\sigma_{B}=\pm\sqrt{\frac{r_{c}-r}{2v_{3}}+|\rho_{1}|\left(\frac{1}{2v_{3}}+\frac{1}{2(v_{3}-v_{1})}\right)},\label{rhoC2}
\end{align}
and in the collinear $C_{4}$ AFM phase,
\begin{align}
\rho_{1} & =0,\rho_{2}=\pm a_{2}e^{W_{0}\left(\frac{r_{c}-r}{4\overline{v}_{3}a_{2}}\right)}\stackrel{r\rightarrow r_{c}^{-}}{\longrightarrow}\pm a_{2},\\
\sigma_{A/B} & =\pm\sqrt{\frac{r_{c}-r}{v_{3}}+|\rho_{2}|\left(\frac{1}{v_{3}}+\frac{1}{v_{3}-v_{1}}\right)},\sigma_{B/A}=0,\label{rhoC4}
\end{align}
where $a_{i}=\frac{c\Lambda^{2}}{2}\exp\left\lbrace 1-\frac{1}{4(\overline{v}_{3}-\overline{v}_{i})}+\frac{1}{4\overline{v}_{3}}\right\rbrace $,
$\overline{v}_{i}=\frac{v_{i}}{4\pi^{2}c\gamma}$, $r_{c}=-2v_{3}I(0)$.
$W_{0}(z)$ is the principal branch of the Lambert function, which
is the solution of $z=We^{W}$. The nonvanishing values of $\rho_1$ and $\rho_2$ approaching to the transition 
indicate it is first-order. From Eqs.~\eqref{rhoC2} and \eqref{rhoC4}, it is obvious that the magnetic transition just follows the one between the composite orders, and is also first-order. 
The free energy density at the saddle-point level can then be expressed 
in terms of the composite order parameters only.
\begin{align}
	f=\frac{S^{(0)}_{eff}}{N\beta V}=\left(\frac{1}{4(v_{3}-v_{1})}-\frac{1}{4v_{3}} \right)\rho_{1}^{2}+ \left(\frac{1}{4(v_{3}-v_{2})}-\frac{1}{4v_{3}} \right)\rho_{2}^{2}+\frac{r}{2v_{3}}\rho-\frac{r^{2}}{4v_{3}}+\frac{1}{2\beta V}\sum_{\bm{q},i\omega_{l}}\left[ \ln{\bar{\chi}_{0}^{-1}}+\ln{(\bar{\chi}_{0}^{-1}+2\rho)}\right],
\end{align}

which gives rise to the new saddle-point equations 
\begin{align}
	\rho_{1}\left( \frac{1}{2(v_{3}-v_{1})}-\frac{1}{2v_{3}}+\frac{r}{2v_{3}\rho}+\frac{1}{\rho}\frac{1}{\beta V}\sum_{\bm{q},i\omega_{l}}\frac{1}{\bar{\chi}_{0}^{-1}+2\rho}\right)=0,\label{nmag1}\\
		\rho_{2}\left( \frac{1}{2(v_{3}-v_{2})}-\frac{1}{2v_{3}}+\frac{r}{2v_{3}\rho}+\frac{1}{\rho}\frac{1}{\beta V}\sum_{\bm{q},i\omega_{l}}\frac{1}{\bar{\chi}_{0}^{-1}+2\rho}\right)=0,\label{nmag2}
\end{align}
under the constraints $m=\rho, \sigma_{A}\sigma_{B}=-\sigma^{2}\rho_{1}/\rho$ and $(\sigma_{A}^{2}-\sigma_{B}^{2})/2=-\sigma^{2}\rho_{2}/\rho$. Note that this free energy form is still analytic since it is in the magnetically ordered phase.

To 
calculate the nematic and charge susceptibilities inside the magnetically ordered phases, we 
consider fluctuations beyond the saddle-point level. 
For the stability of the theory, we 
implement the constraint $m=\rho$ 
even beyond the saddle point. 
By separating the static (saddle-point) and fluctuating parts of the order parameter fields, $\rho_{i}\rightarrow\rho_{i}+\phi_{i}$, $\rho\rightarrow\rho +\rho_{1}\phi_{1}/\rho +\rho_{2}\phi_{2}/\rho +\rho_{2}^{2}\phi_{1}^{2}/2\rho^{3} +\rho_{1}^{2}\phi^{2}_{2}/2\rho^{3}$, we obtain
\begin{align}	S_{eff}^{(2)}=\frac{1}{2}\sum_{\bm{q},i\omega_{l}}\chi_{nem}^{-1}(\bm{q},i\omega_{l})|\phi_{1}(\bm{q},i\omega_{l})|^{2}+\chi_{cha}^{-1}(\bm{q},i\omega_{l})|\phi_{2}(\bm{q},i\omega_{l})|^{2}\label{field theory3}
\end{align}
with the nematic/charge $C_4$ susceptibilities
\begin{align}
\chi_{nem/cha}^{-1}(\bm{q},i\omega_{l}) =\frac{1}{2(v_{3}-v_{1/2})}-\frac{1}{2v_{3}} +\frac{r}{2v_{3}}\frac{\rho_{2/1}^{2}}{\rho^{3}}-\pi_{1/2}(\bm{q},i\omega_{l}),
\end{align}
where
\begin{align}
\pi_{1/2}(\bm{q},i\omega_{l})=&\frac{2}{\beta V}\sum_{\bm{k},i\omega_{n}}\frac{1}{(\bar{\chi}_{0}^{-1}(\bm{k},i\omega_{n})+2\rho)(\bar{\chi}_{0}^{-1}(\bm{k}+\bm{q},i\omega_{n}+i\omega_{l})+2\rho)}\frac{\rho^{2}_{1/2}}{\rho^{2}}\nonumber\\
&-\frac{1}{2\beta V}\sum_{\bm{k},i\omega_{n}}\left(\frac{1}{\bar{\chi}_{0}^{-1}(\bm{k},i\omega_{n})}
+\frac{1}{\bar{\chi}_{0}^{-1}(\bm{k},i\omega_{n})+2\rho} \right)\frac{\rho_{2/1}^{2}}{\rho^{3}} \nonumber\\
&+\frac{1}{\beta V}\sum\limits _{\bm{k},i\omega_{n}}\frac{(\bar{\chi}_{0}^{-1}(\bm{k},i\omega_{n})+\rho)(\bar{\chi}_{0}^{-1}(\bm{k}+\bm{q},i\omega_{n}+i\omega_{l})+\rho)-\rho^{2}}{[(\bar{\chi}_{0}^{-1}(\bm{k},i\omega_{n})+\rho)^{2}-\rho^{2}][(\bar{\chi}_{0}^{-1}(\bm{k}+\bm{q},i\omega_{n}+i\omega_{l})+\rho)^{2}-\rho^{2}]}\frac{\rho_{2/1}^{2}}{\rho^{2}}\nonumber\\
=&\frac{2}{\beta V}\sum_{\bm{k},i\omega_{n}}\frac{1}{(\bar{\chi}_{0}^{-1}(\bm{k},i\omega_{n})+2\rho)(\bar{\chi}_{0}^{-1}(\bm{k},i\omega_{n})+2\rho)}\frac{\rho^{2}_{1/2}}{\rho^{2}}-\frac{1}{\beta V}\sum_{\bm{k},i\omega_{n}}\frac{1}{\bar{\chi}_{0}^{-1}(\bm{k},i\omega_{n})+2\rho} \frac{\rho_{2/1}^{2}}{\rho^{3}}\nonumber\\
&+\pi_{1/2}(\bm{q},i\omega_{l})-\pi_{1/2}(0,0).
\end{align}
In $C_{2}$ collinear(nematic) phase, $\rho_{1}\neq 0$ and $\rho_{2}=0$, using Eqs.(\ref{nmag1},\ref{nmag2}),
\begin{align}
\chi_{nem}^{-1}(\bm{q},i\omega_{l})=&\frac{1}{2(v_{3}-v_{1})}-\frac{1}{2v_{3}}-\frac{2}{\beta V}\sum_{\bm{k},i\omega_{n}}\frac{1}{(\bar{\chi}_{0}^{-1}(\bm{k},i\omega_{n})+2\rho)(\bar{\chi}_{0}^{-1}(\bm{k}+\bm{q},i\omega_{n}+i\omega_{l})+2\rho)}\\
\chi_{cha}^{-1}(\bm{q},i\omega_{l})=&\frac{1}{2(v_{3}-v_{2})}-\frac{1}{2v_{3}}+\frac{r}{2v_{3}}\frac{\rho_{1}^{2}}{\rho^{3}}+\frac{1}{\rho}\frac{1}{\beta V}\sum_{\bm{k},i\omega_{n}}\frac{1}{\bar{\chi}_{0}^{-1}(\bm{k},i\omega_{n})+2\rho}+\pi_{2}(0,0)- \pi_{2}(\bm{q},i\omega_{l})\nonumber\\
=&\frac{1}{2(v_{3}-v_{2})}-\frac{1}{2(v_{3}-v_{1})}+\pi_{2}(0,0)- \pi_{2}(\bm{q},i\omega_{l})
\end{align}
and in $C_{4}$ collinear(charge) phase, $\rho_{1}= 0$ and $\rho_{2}\neq0$,
\begin{align}
	\chi_{cha}^{-1}(\bm{q},i\omega_{l})=&\frac{1}{2(v_{3}-v_{2})}-\frac{1}{2v_{3}}-\frac{2}{\beta V}\sum_{\bm{k},i\omega_{n}}\frac{1}{(\bar{\chi}_{0}^{-1}(\bm{k},i\omega_{n})+2\rho)(\bar{\chi}_{0}^{-1}(\bm{k}+\bm{q},i\omega_{n}+i\omega_{l})+2\rho)}\\
	\chi_{nem}^{-1}(\bm{q},i\omega_{l})=&\frac{1}{2(v_{3}-v_{1})}-\frac{1}{2v_{3}}+\frac{r}{2v_{3}}\frac{\rho_{2}^{2}}{\rho^{3}}+\frac{1}{\rho}\frac{1}{\beta V}\sum_{\bm{k},i\omega_{n}}\frac{1}{\bar{\chi}_{0}^{-1}(\bm{k},i\omega_{n})+2\rho}+\pi_{1}(0,0)- \pi_{1}(\bm{q},i\omega_{l})\nonumber\\
	=&\frac{1}{2(v_{3}-v_{1})}-\frac{1}{2(v_{3}-v_{2})}+\pi_{1}(0,0)- \pi_{1}(\bm{q},i\omega_{l})
\end{align}

It is then easy to see 
that approaching to the transition as $v_{1}\rightarrow v_{2}$, $\chi_{nem/cha}$ diverges in the 
double-Q/single-Q AFM phase, signifying the emergence of the Goldstone mode.

\subsection{The effect of the spin anisotropy}
In the presence of anisotropic spin fluctuations described by the $w\neq 0$ term in Eq.~(1) of the main text, the emergent $U(1)$ symmetry is only approximate, and the Goldstone mode is gapped out. 
However, for small anisotropy $w/c\ll1$, there should still be a pseudo-Goldstone mode.
In this section, 
we show that the gapless Goldstone mode is robust 
up to 
the $(w/c)^{3}$ order compared with the gapped Higgs mode, which means that the ratio between the 
pseudo-Goldstone gap and the Higgs gap is of (or beyond) the 
$(w/c)^{4}$ order.

We still focus on the nematic to 
charge $C_4$ transition in the paramagnetic phase. By 
including the anisotropy, Eq.(\ref{matrix}) becomes:
\begin{flalign}
	tr\ln{\bm{D}^{-1}}=\ln{det{\begin{pmatrix}\bar{\chi}_{0}^{-1}(x,\tau)+i\lambda+\rho_{2} & \rho_{1}-w(\partial_{x}^{2}-\partial_{y}^{2})+\bm{\rho}_{3}\\
				\rho_{1}-w(\partial_{x}^{2}-\partial_{y}^{2})-\bm{\rho}_{3} & \bar{\chi}_{0}^{-1}(x,\tau)+i\lambda-\rho_{2}
	\end{pmatrix}}}.\label{matrix2}
\end{flalign}
For small anisotropy $w/c\ll 1$, we 
can expand the action in powers of $w/c$ to the quadratic order,
\begin{align}
	S_{w}=S_{eff}-\left(\frac{w}{c} \right)^{2}\frac{1}{\beta V}\sum_{\bm{q},i\omega_{n}}\frac{c^{2}(q_{x}^{2}-q_{y}^{2})^{2}\left[(m+cq^{2}++\gamma|\omega_{n}|)^{2}+\rho_{1}^{2}-\rho_{2}^{2} \right] }{\left[(m+cq^{2}+\gamma|\omega_{l}|)^{2}-\rho^{2} \right]^{2}}.
\end{align}
After some simplifications, and assuming weakly first-order transition $(\rho\ll m)$, the action involving the composite order
fields has the following form:
\begin{align}	S_{w}=\frac{1}{2}\sum_{\bm{q},i\omega_{l}}\sum_{i=1,2}(r_{i}+K_{x}q^{2}+K_{\tau}\omega_{l}^{2})|\rho_{i}(\bm{q},i\omega_{l})|^{2}+g\int_{0}^{\beta}d\tau\int d^{2}x (\rho_{1}^{2}+\rho_{2}^{2})^{2}
-\epsilon_{1}^{2}(\rho_{1}^{4}-\rho_{2}^{4})-\epsilon_{2}^{4}(\rho_{1}^{4}+\rho_{2}^{4}),
\end{align}
where \begin{align}
	\epsilon_{1}^{2}=&\frac{1}{g} \left( \frac{w}{c}\right)^{2}\frac{1}{\beta V}\sum_{\bm{q},\omega_{l}}\frac{c^{2}q^{4}}{(m+cq^{2}+\gamma|\omega_{l}|)^{6}}
	=\frac{1}{g} \left( \frac{w}{c}\right)^{2}\frac{1}{120(2\pi)^{2}c\gamma m^{2}}
	\\
	\epsilon_{2}^{4}=&\frac{3}{8g}\left(\frac{w}{c} \right)^{4}\frac{1}{\beta V}\sum_{\bm{q},\omega_{l}}\frac{c^{4}q^{8}}{(m+cq^{2}+\gamma|\omega_{l}|)^{8}}=\frac{1}{g}\left(\frac{w}{c} \right)^{4}\frac{1}{560(2\pi)^{2}c\gamma m^{2}} ,
	\end{align}
and $r_{1},r_{2},g$ are correspondingly shifted from their isotropic values. 
In this case, the transition between nematic and charge order occurs at
\begin{align}
	\frac{r_{1}^{2}}{g(1-\epsilon_{1}^{2}-\epsilon_{2}^{4})}=\frac{r_{2}^{2}}{g(1+\epsilon_{1}^{2}-\epsilon_{2}^{4})}.
\end{align}
This means that the original transition point ($r_1/r_2=1$) is shifted by the order of $O\left(\frac{w}{c}\right)^2$.

The
mass (gap) of the nematic fluctuations 
approaching to the new transition point from the charge $C_4$
phase is then 
\begin{align}
	\delta_{nem}=&r_{1}-\frac{r_{2}}{1+\epsilon_{1}^{2}-\epsilon_{2}^{4}}\nonumber\\
	=&\frac{\delta_{cha}}{2}\left(\frac{1}{2}\epsilon_{1}^{4}+\epsilon_{2}^{4} \right)
	,
\end{align}
where $\delta_{cha}=-2r_{2}$. Note that the gap $\delta_{nem}/\delta_{cha}\sim O\left(\frac{w}{c}\right)^4$.
The same conclusion applies to the $C_4$ charge fluctuations in the nematic phase.

\subsection{Thermodynamic signature of the emergent symmetry}

Because of the divergence of the correlation lengths, we expect to observe quantum critical behavior around the first-order transition. 
The
singular part of the free energy 
can be written in terms of the nematic and charge masses 
$\delta_{nem}$ and $\delta_{cha}$. At 
temperature $T$,
\begin{align}
f_{s}(T,g)\propto T^{(d+z)/z}\Phi\left(\frac{\delta_{nem}}{T^{1/z\nu}},\frac{\delta_{cha}}{T^{1/z\nu}}\right).
\end{align}
where
$\delta_{nem}=(g-g_{c})/g_{c}$
in the charge $C_4$ phase, $\delta_{cha}=(g_{c}-g)/g_{c}$ in the nematic phase. Here $g$ is a tuning parameter and $g_c$ denotes the transition point. Though the scaling function contains two arguments, 
approaching the transition point, only one mass can be tuned to zero. As a result, the Gr\"{u}neisen ratio $\Gamma=\alpha/c_p$
scales as 
\begin{align}
\Gamma_{cr}(T,g \rightarrow g_{c}^{\pm})\sim \pm T^{-\frac{1}{\nu z}},
\end{align}
taking the same form as near a QCP \cite{zhu2003}. 

Note again that the anisotropic XZ model,  shares similarities with our model. It has  been shown to exhibit similar divergent behavior in the Gr\"{u}neisen ratio \cite{Beneke2021}.

\subsection{Theory of Phonon Softening across Nematic Transition}
In this section we propose that the in-plane transverse phonons could be used to probe the nematic fluctuations in different phases. We are motivated by a recent experiment on the NaFeAs compound, which reveals pronounced in-plane transverse
acoustic phonon softening along Fe-As direction across $T_{s}$ where
the nematic order is developed\cite{Li_PRX:2018}.

Since $B_{1g}$ nematicity transforms as $x^{2}-y^{2}$ under point group transformations, the nematoelastic coupling has the following form due to symmetry considerations:

\begin{align}
    H_{nem-el}=\lambda\int _{x}\rho_{1}(\partial_{x}u_{x}-\partial_{y}u_{y})
\end{align}
where $\lambda$ is the coupling constant, and $\bm{u}$ is the atomic displacement vector which could be quantized to phonons:
\begin{align}
    \bm{u}({\bm{R}_{i}})=-\frac{i}{\sqrt{2MN\omega_{\bm{q}s}}}\sum_{\bm{q}s}\hat{\bm{e}}_{\bm{q}s}e^{-i\bm{q}\cdot\bm{R}_{i}}(b_{\bm{q}s}+b^{\dagger}_{-\bm{q}s})
\end{align}
where $\omega_{\bm{q}s}=c_{s}q$ is the energy of the acoustic phonon with velocity $c_{s}$ at polarization direction $s$,  and $M$ is the effective ion mass. For the in-plane transverse branch of the phonons ($s=T$), $\hat{\bm{e}}_{\bm{q}T
}=\hat{z}\times \hat{\bm{q}}=(-q_{y}/q,q_{x}/q,0)$, the nematoelastic coupling in this branch becomes
\begin{align}
    H_{nem-el}=g_{T}\sum_{\mathbf{q}}\frac{q_{x}q_{y}}{q\sqrt{q}}\rho_{1}(\bm{q})(b_{\mathbf{q}T}+b_{-\mathbf{q}T}^{\dagger}), \label{couple}
\end{align}
with $g_{T}=\frac{2\lambda}{\sqrt{2Mc_{s}}}$. Combining Eq.(\ref{field theory2}) with Eqs.(\ref{couple}),  we get a renormalized nematic fluctuation by integrating out phonons,
\begin{align}
	\tilde{\chi}_{1}^{-1}(q,i\omega_{n})=&r_{1}+Kq^{2}+\omega_{n}^{2}-\frac{2g_{1T}^{2}\omega_{\mathbf{q}}}{\omega_{n}^{2}+\omega_{\mathbf{q}}^{2}}\frac{q_{x}^{2}q_{y}^{2}}{q^{3}}
\end{align}

As the Gaussian part nematic fluctuation field $\phi_{1}$ becomes critical,
the Ising nematic phase transition takes place, where critical value
$r_{1c}$ satisfies the condition: 
\begin{equation}
	\min_{q,\omega_{n}}\left[r_{1c}+Kq^{2}+\omega_{n}^{2}-\frac{2g_{1T}^{2}\omega_{\mathbf{q}}}{\omega_{n}^{2}+\omega_{\mathbf{q}}^{2}}\frac{q_{x}^{2}q_{y}^{2}}{q^{3}}\right]=0,
	\label{stability}\end{equation}
From the above equation we get the critical value $r_{1c}=g_{1T}^{2}/2c_{T}$, 
where the minimum is reached at $\omega_{n}=0$ and $q_{x}=q_{y}=q/\sqrt{2}\rightarrow0$,
precisely correspond to the Fe-As direction where phonon softening was
observed. 

The coupling between acoustic phonons and nematic fluctuations should
correct the energy of both modes. Integrating out the fluctuations / phonons, and making $i\omega_{n}\rightarrow \omega+0^{+}$, we get the renormalized energies for the nematic / phonon modes:
\begin{align}
	\omega_{\pm}^{2}=\frac{r_{1c}}{2}\left\lbrace  f(\delta,\tilde{q},\epsilon)\pm\sqrt{f(\delta,\tilde{q},-\epsilon)^{2}+4\epsilon \tilde{q}^{2}}\right\rbrace  \label{phe}
\end{align}
with
\begin{align}
	f(\delta,\tilde{q},\epsilon)=1+\delta+(1+\epsilon)\tilde{q}^{2}
\end{align}
where $\omega_{\pm}$ represents Ising nematic mode (+) and phonon mode (-), respectively; $\delta=(r_{1}-r_{1c}+12g\overline{\rho}_{1}^{2}+4g\overline{\rho}_{2}^{2})/r_{1c}$; $\epsilon=c_{T}^{2}/K$; $\tilde{q}=\sqrt{K/r_{1c}}q$; $r_{1}+12g\overline{\rho}_{1}^{2}+4g\overline{\rho}_{2}^{2}$  denotes for the square of the gap of the nematic fluctuations, where $\overline{\rho}_{1}$ and $\overline{\rho}_{2}$ are the nematic order parameter and the charge order parameter at the saddle point level. We find that the spin-phonon coupling
leads to repulsion of phonon and nematic energy levels, resulting in phonon softening
and nematicity hardening. At the nematic critical point $(\delta=0)$, where the nematic suseptibility diverges, the phonon dispersion $\omega_{-}(q\rightarrow 0)\rightarrow \sqrt{r_{1c}\epsilon}\tilde{q}^{2}$, which indicates the phonon dispersion is softened from linear to quadratic. Meanwhile, the coupling to the gapless phonon mode opens a
gap to the nematic mode, $\omega_{+}(q\rightarrow 0)\rightarrow\sqrt{r_{1c}}$.

Since the IPTA phonons are sensitive to the nematicity, they can be used to detect the nematic fluctuations in different phases.  An observable quantity is the phonon velocity:
\begin{align}
	\tilde{c}_{T}=\lim_{q\rightarrow 0}\frac{\omega_{-}(q)}{q}=c_{T}\sqrt{\frac{\delta}{1+\delta}}
\end{align}
From disordered phase to the nematic phase, the transition is continuous and the phonon velocity gradually vanishes as the nematic transition is reached, then it will gradually increase after the transition. On the other hand, from charge order to nematic order, where the transition is first order, the phonon velocity gradually vanishes as well which represents the emergence of the Goldstone mode, but will suddenly back to finite value after entering into nematic phase, as illustrated in Fig.\ref{fig3}.

\begin{figure}[tb!]
\centering \includegraphics[width=.6\linewidth]{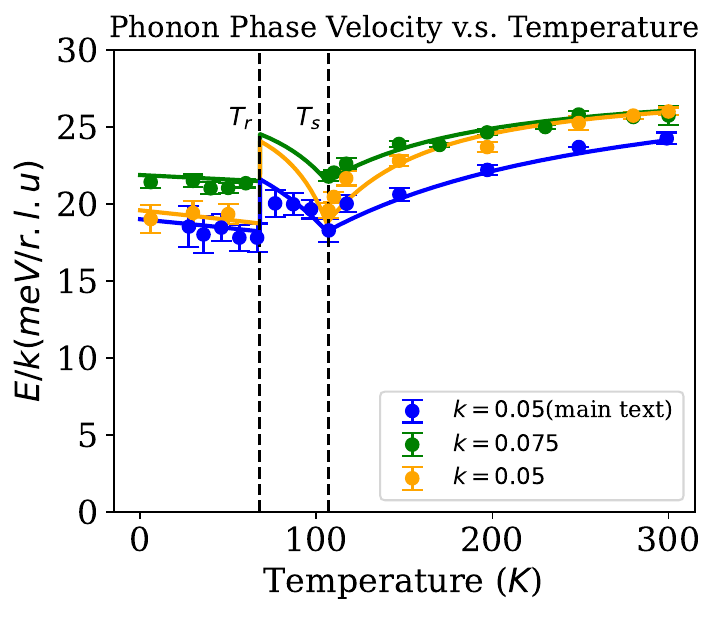} 
\caption{\label{figS1}
Measured and fitted phase velocity $E/k$ of the IPTA phonons 
vs. temperature for Sr$_{0.64}$Na$_{0.36}$Fe$_2$As$_2$ \cite{Song_PRL:2021} at $k=0.05 \, r.l.u$ (solid symbols) and fits with our analytical results (solid line), where the regions $T_{c}<T<T_{r}$, $T_{r}<T<T_{s}$ and $T>T_{s}$ correspond to double-Q AFM, single-Q AFM and tetragonal PM phases respectively. The blue data corresponds to that presented in the main text, while the green and orange data are extracted from Ref. \cite{Song_PRL:2021}.}
\end{figure}

\begin{figure}[tb!]
\centering \includegraphics[width=.6\linewidth]{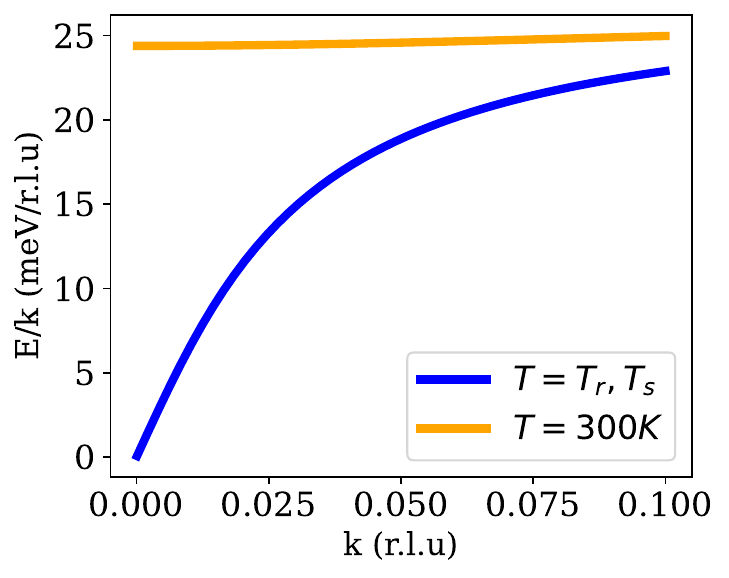} \caption{\label{figS2}
Phase velocity $E/k$ of the IPTA phonons 
vs. temperature for Sr$_{0.64}$Na$_{0.36}$Fe$_2$As$_2$ \cite{Song_PRL:2021} at $T=300K$ (orange) and at $T=T_{r}$ and $T_{s}$ (blue). }
\end{figure}

To align with the observed phonon energy of the in-plane transverse acoustic phonons in Sr$_{0.64}$Na$_{0.36}$Fe$_2$As$_2$ \cite{Song_PRL:2021}, we employed our analytic expressions as outlined in Eq.(\ref{phe}) for fitting the data. We defined $r_{1}/r_{1c}=a(T-T_{s})$ and $r_{2}/r_{1c}=b(T-T_{ch})$, with the difference $(r_{1}-r_{2})/r_{1c}=(b-a)(T_{r}-T)$ established, where $T_{ch}=T_{r}+(T_{s}-T_{r})\frac{a}{b}, T_{s}=107K, T_{r}=68K$. Additionally, we introduced $\tilde{q}=q_{rlu}\xi$, with $q_{rlu}$ representing the momentum in reciprocal lattice units (r.l.u.), where experimentally it is measured at $q_{rlu}=0.075\,r.l.u$ and $q_{rlu}= 0.05 \,r.l.u$. The parameters chosen for the fit include:$a=0.027/K,,b=0.033/K, ,\epsilon=0.77, r_{1c}=2.36 meV, ,\xi=25.7/r.l.u.$. The resulting fitted plot is depicted in Fig.\ref{fig4}  and Fig.\ref{figS1}. We also plot the phonon phase velocity $E/k$ v.s $k$ in Fig.\ref{figS2} for $T=300K$ and $T=T_{r},T_{s}$, where it is clear that phonons are strongly softened for small $k$ both at nematic phase transition $T=T_{s}$ and $C_2-C_{4}$ transition $T=T_{r}$.

We also fitted the data from Ref. \cite{Song_PRL:2021} in Fig.\ref{figS1}. Here, the parameters are  $a=0.07/K,,b=0.075/K, ,\epsilon=0.59, r_{1c}=2.58 meV, ,\xi=26/r.l.u.$. The discrepancy between the two sets of data at $k=0.05\,r.l.u$ arises from different beamlines and measurements (different sample spots). Nonetheless, they reach qualitative agreement in the low temperature regime.

\end{document}